\documentclass{gmp2012}

\usepackage{placeins}
\usepackage{subfigure}
\usepackage{algorithm}
\usepackage{algorithmic}
\usepackage{multirow}

\begin{document}

%
%
\title{Fast B-spline Curve Fitting by L-BFGS}

%
%
\SubNumber{31}

%
%
\author{Wenni Zheng}{1}
\author{Pengbo Bo}{1, 2}
\author{Yang Liu}{3}
\author{Wenping Wang}{1}

%
%
\affiliation{1}{The University of Hong Kong}
\affiliation{2}{Harbin Institute of Technology at Weihai}
\affiliation{3}{Microsoft Research Asia}
%
%


\maketitle

\begin{abstract}

We propose a novel method for fitting planar B-spline curves to unorganized data points. In traditional methods, optimization of control points and foot points are performed in two very time-consuming steps in each iteration: 1) control points are updated by setting up and solving a linear system of equations; and 2) foot points are computed by projecting each data point onto a B-spline curve. Our method uses the L-BFGS optimization method to optimize control points and foot points simultaneously and therefore it does not need to perform either matrix computation or foot point projection in every iteration. As a result, our method is much faster than existing methods.

\end{abstract}


\section{Introduction}

Curve fitting is a fundamental problem in many fields, such as computer graphics, image processing, shape modeling and data mining.  Depending on applications, different types of curves such as parametric curves, implicit curves and subdivision curves are used for fitting. In this paper, we study the problem of fitting planar B-spline curves to unorganized data points.

Given a set of unorganized data points $\{\mathbf{X}_i\}_{i=1}^N\subset \mathbb{R}^2$ sampled from the outline of a planar shape, the aim of curve fitting is to find a B-spline curve $\mathbf{P}(t)=\sum_{i=1}^n\mathbf{P}_iN_i(t)$ that best approximates the shape's outline. The outline is called a \emph{target shape}, and the B-spline curve is called a \emph{fitting curve}. Here, $\mathcal{P}:= \{\mathbf{P}_i\}_{i=1}^n \subset \mathbb{R}^2$ is the set of B-spline control points, $\{N_i(t)\}_{i=1}^n$ are B-spline basis functions. We suppose that knots of the B-spline curve are fixed and therefore not subject to optimization, and all the basis functions are thus defined on fixed, uniform spaced knots throughout the curve fitting process.

For a data point $\mathbf{X}_k$, let $\mathbf{P}(t_k)$ denote the nearest point of $\mathbf{X}_k$ on the fitting curve. Then, the distance between data point $\mathbf{X}_k$ and the fitting curve is $ \|\mathbf{P}(t_k)-\mathbf{X}_k\|$. Here, $t_k$ is called the \emph{location parameter} of $\mathbf{X}_k$, $\mathbf{P}(t_k)$ is called the \emph{foot point} corresponding to $\mathbf{X_k}$. Denote $\mathcal{T}=\{ t_1,...,t_k \}$, i.e. the collection of the location parameters of all the data points.
The fitting problem is then formulated as:
\begin{equation}\label{eq:general}\min\limits_{\mathcal{P},\mathcal{T}} \frac{1}{2}\sum\limits_{k=1}^N \|\mathbf{P}(\mathcal{P};t_k)-\mathbf{X}_k\|^2+\lambda F_{fairing},\end{equation}
where $F_{fairing}$ is a fairing term which defines the fairness of a curve. $F_{fairing}$ is commonly defined as follows~\cite{Wang2006}:
\begin{equation}\label{eq:oldfairing}
F_{fairing} = \alpha\int_0^1\|\mathbf{P}'(t)\|^2dt+\beta\int_0^1\|\mathbf{P}''(t)\|^2dt.
\end{equation}

Since the objective function in Eqn.~\ref{eq:general} is nonlinear, it is natural to apply iterative minimization methods to solve it.  Most prevailing methods for solving this problem in CAGD are not standard optimization methods in the sense that they separately optimize location parameters $\mathcal{T}$ and control points $\mathcal{P}$, making the problem much simpler to handle. However, these methods are time-consuming because they need to compute foot points on the fitting curve and to formulate and solve linear systems in every iteration. We observe that these time-consuming operations can be avoided by employing a L-BFGS optimization method that solves $\mathcal{T}$ and $\mathcal{P}$ simultaneously. We show that the resulting algorithm is very efficient because in every iteration it does not need to perform foot point projection or solve a linear system of equations.

The remainder of this paper is organized as follows. In section 2, we review some previous work. Section 3 introduces the standard L-BFGS optimization method.  Section 4 presents our new algorithm. Section 5 shows experimental results and comparisons with existing methods. Then we conclude the paper in Section 6 with discussions of future work.

\section{Related Work}

Problem~\ref{eq:general} can also be formulated as a nonlinear constrained minimization problem with unknown variables $\mathcal{P}$:
\begin{equation}\label{eq:general_const}\min\limits_\mathcal{P} \frac{1}{2}\sum\limits_{k=1}^N \|\mathbf{P}(t_k)-\mathbf{X}_k\|^2+\lambda F_{fairing},\end{equation}
where each $t_k$ is chosen such that $\mathbf{P}(t_k)$ is the foot point of $\mathbf{X}_k$ and thus satisfies:
\begin{equation}\label{eq:foot} (\mathbf{P}(t_k)-\mathbf{X}_k)^T\mathbf{P}'_t(t_k)=0,\qquad k=1,\ldots,N.\end{equation}
By representing $\{t_k\}$ as functions of $\mathcal{P}$ and putting them into the objective function in Eqn.~\ref{eq:general_const}, we obtain a nonlinear unconstrained minimization problem of variables $\mathcal{P}$:
\begin{equation}\label{eq:global}
\min\limits_{\mathcal{P}} \frac{1}{2}\sum\limits_{k=1}^N \|\mathbf{P}(t_k(\mathcal{P}))-\mathbf{X}_k\|^2+\lambda F_{fairing}.
\end{equation}
This is the viewpoint taken in~\cite{Liu2008} that reveals inherent relationship between some traditional methods and standard optimization techniques. Most methods for solving problem~\ref{eq:global} deal with control points and foot points separately~\cite{Hoschek:1989:OAC:80732.80734}~\cite{Liu2008}. Each iteration of these methods consists of the following two steps:

{\bf Step 1}: \emph{Foot point projection}: Fixing the control points of the current fitting curve, compute the location parameters $\mathcal{T} = \{t_k\}$ for the data points $\{\mathbf{X}_k\}$ such that $\{\mathbf{P}(t_k)\}$ are the foot points of $\{\mathbf{X}_k\}$ on the current fitting curve. This step preserves the orthogonality constraint in Eqn.~\ref{eq:foot}.

{\bf Step 2}: \emph{Control point update}: In this step, $\mathcal{T}$ is fixed and a quadratic function $e_{k}$ in terms of the control points $\mathcal{P}$ is used to approximate the nonlinear squared distance from a data point $\mathbf{X}_k$ to the fitting curve. Then the control points $\mathcal{P}$ are computed by minimizing the quadratic objective function $Q(\mathcal{P}) = \sum_{k}e_{k}(\mathcal{P}) + \lambda F_{fairing}$. Since both $e_{k}$ and  $F_{fairing}$  are quadratic functions of $\mathcal{P}$, this step entails solving the linear equations $\nabla Q(\mathcal{P}) = 0$.

Depending on different quadratic approximations chosen for $e_k$, there are mainly three kinds of existing optimization methods for curve fitting. The first one is the \emph{Point Distance Minimization} method, or PDM. This method is widely used because of its simplicity. References on PDM include (but are not limited to)~\cite{Pavlidis1983},~\cite{Plass:1983:CPP:800059.801153},~\cite{Hoschek1988} and~\cite{Saux:2003:IHI:965907.965910} on curve fitting as well as~\cite{Hoppe:1994:PSS:192161.192233},~\cite{Forsey:1995:SFH:221659.221665} and~\cite{haber2001smooth} on surface fitting. 
The error term used in PDM is defined by
\begin{equation}
e_{PD,k} = \|\mathbf{P}(\mathcal{P};t_k)-\mathbf{X}_k\|^2.
\end{equation}
Geometrically, this function defines the distance between a data point and a point on the fitting curve at a particular parameter $t_k$. Considering the fact that $t_k(\mathcal{P})$ is set to a constant, this definition is a poor approximation of the nonlinear distance in Eqn.~\ref{eq:global}. As pointed out in~\cite{Bjorck1996}, from the viewpoint of optimization, PDM is an alternating method and exhibits linear convergence rate. We will see in our experiments that PDM is the slowest among all the methods we have tested.

The second method is called the \emph{tangent distance minimization} method (TDM)~\cite{Blake1998} which uses the error term
\begin{equation}
e_{TD,k} = \left[(\mathbf{P}(\mathcal{P};t_k)-\mathbf{X}_k)^T\cdot \mathbf{N}_k \right]^2,
\end{equation}
where $\mathbf{N}_k$ is the unit normal vector at point $\mathbf{P}(t_k)$ on the curve.

The term $e_{TD,k}$ defines the distance between a data point $\mathbf{X}_k$ and the tangent line at $\mathbf{P}(t_k)$. Although this is a fair approximation to the true squared distance near a flat part of curve, it is not accurate near high curvature regions since no curvature information is considered. As a result, TDM does not show stable performance near high curvature regions~\cite{Wang2006}. In fact, it has been pointed out in~\cite{Wang2006} that TDM is essentially Gauss-Newton minimization without step-size control, and regularization should be used to improve the stability of TDM.

Applying the Levenberg-Marquardt regularization to TDM leads to a method called TDMLM~\cite{Wang2006}.  Suppose the linear system for control points updating in TDM is $A_{TDM}\cdot\mathcal{P}=\mathbf{b}_{TDM}$, where $A_{TDM}$ is a matrix and $\mathbf{b}_{TDM}$ is a vector. In TDMLM,  the control points $\mathcal{P}$ are computed by solving
\[(A_{TDM}+\mu I)\cdot\mathcal{P}=\mathbf{b}_{TDM}.\]  
Empirically, $\mu$ is set as $\mu=\frac{tr(A_{TDM})}{80n}$, where $tr(A_{TDM})$ is the trace of $A_{TDM}$, $n$ the number of control points, and $I$ the identity matrix.

The third method, called the \emph{Squared Distance Minimization} method or SDM~\cite{Wang2006}, uses a curvature-based error term, which is a variant of the second order approximation to the true squared distance introduced in~\cite{Pottmann2002}~\cite{Pottmann2003}. This error term, called the {\em SD error term}, is defined by
\begin{equation}
e_{SD,k}=\left\{\begin{array}{l}\frac{d}{d-\rho}\left[(\mathbf{P}(\mathcal{P};t_k)-\mathbf{X}_k)^T\cdot T_k\right]^2+\\
\quad+\left[(\mathbf{P}(\mathcal{P};t_k)-\mathbf{X}_k)^T\cdot N_k\right]^2,\mbox{if }d<0,\\
\left[(\mathbf{P}(\mathcal{P};t_k)-\mathbf{X}_k)^T\cdot N_k\right]^2,\mbox{if }0\le d<\rho,\end{array}\right.
\end{equation}
where $\rho$ is the curvature radius at $\mathbf{P}(t_k)$ and $d$ is the positive distance between $\mathbf{X}_k$ and $\mathbf{P}(t_k)$. The SD error term contains some second order derivative information and is therefore a better approximation to the true squared distance function than those used in TDM and PDM. From the viewpoint of optimization, SDM is quasi-Newton optimization method that employs a modified Hessian matrix of the original nonlinear distance function. This modification discards some complicated parts in the true Hessian matrix and keeps other parts with intuitive geometric meanings~\cite{Wang2006}. The semi-definite positive property of the modified Hessian matrix is also guaranteed. It has been demonstrated in~\cite{Wang2006} that SDM exhibits better performance in terms of convergence rate and stability than PDM and TDM.

Since the curve fitting problem is formulated as a nonlinear least squares minimization problem in Eqn.~\ref{eq:general}, it is natural to study how to solve it using standard optimization methods. The authors of~\cite{Liu2008} apply the Gauss-Newton method to Eqn.~\ref{eq:general} and derive new error terms using simplified partial derivatives of the objective function in Eqn.~\ref{eq:general}. These methods are observed to have similar performances as SDM.

All the above methods update control points $\mathcal{P}$ and location parameters $\{t_k\}$  in two interleaving steps. The main difference of our new method with these existing methods is that in every iteration we update $\mathcal{P}$ and $\{t_k\}$ simultaneously. In this sense the most closely related work is~\cite{Speer1998} which also optimizes control points and location parameters simultaneously in every iteration. However, that method uses the Gauss-Newton optimization and therefore still needs to valuate and store the Jacobian matrices of the objective function, whose size depends on the number of data points and control points~\cite{Speer1998}, as well as to solve a linear system of equations.  In contrast, our approach based on L-BFGS does not need to formulate and solve any linear equations and is therefore faster than the method in~\cite{Speer1998}, as we are going to demonstrate in later experiments.

Other optimization techniques have been explored for surface and curve fitting problems in literature. The authors of ~\cite{xie2001automatic} proposed a method for NURBS curve and surface fitting which optimizes control points, parameters and knots by a conjugate gradient method. Genetic Algorithms and optimal control methods have also been tried in curve fitting~\cite{Sarfraz2006494}~\cite{Alhanaty2001a}. These methods are generally slow and have only been applied to simple examples. 

\section{Limited Memory BFGS -- L-BFGS}

Limited Memory BFGS, or L-BFGS, is a quasi-Newton method for solving unconstrained nonlinear minimization problems~\cite{Nocedal1999}. L-BFGS approximates the inverse Hessian matrix of the objective function by a sequence of gradient vectors from previous iterations. Suppose we want to solve an unconstrained optimization problem
\[\min\limits_{x}f(x),\]
where $f(x)$ is a nonlinear function to minimize and $x$ a set of unknown variables. In the $k$-th iteration of L-BFGS, the variables $x_{k+1}$ are updated by
\[\label{eq:newton}x_{k+1}=x_k-\alpha_k H_k \nabla f(x_k),\]
where $H_k$ is an approximation to the inverse Hessian matrix of $f(x)$ at $x_k$. Here, $-H_k \nabla f(x_k)$ is a search direction, and $\alpha_k$ a scalar variable controlling the step-size of search direction~\cite{Nocedal1999}.

Define $s_k:=x_{k+1}-x_k$, $y_k:=\nabla f_{k+1}-\nabla f_k$,  $\rho_k=\frac{1}{y_k^Ts_k}$, $V_k=I-\rho_ky_ks_k^T$. L-BFGS uses  the values of the objective function and its gradient in the $(k-m)$-th iteration through $(k-1)$-th iteration to compute $H_k$~\cite{Nocedal1999}:
\begin{equation}
\setlength\arraycolsep{0.1em}
\begin{array}{rl}
H_k=&(V_{k-1}^T\cdots V_{k-m}^T)H_k^0(V_{k-m}\cdots V_{k-1})\\
&+\rho_{k-m}(V _{k-1}^T\cdots V_{k-m+1}^T)s_{k-m}s_{k-m}^T(V_{k-m+1}\cdots V_{k-1})\\
&+\rho_{k-m+1}(V_{k-1}^T\cdots V_{k-m+2}^T)s_{k-m+1}\cdot\\
&\cdot s_{k-m+1}^T(V_{k-m+2}\cdots V_{k-1})\\
&+\cdots\\
&+\rho_{k-1}s_{k-1}s_{k-1}^T,
\end{array}
\end{equation}
where $H_k^0$ is a diagonal matrix defined by $H_k^0 = \gamma_k I$, where $\gamma_k =\frac{s_{k-1}^Ty_{k-1}}{y_{k-1}^Ty_{k-1}}$~\cite{Nocedal1999}.

In practice, we do not need to compute and store the matrix $H_k$. Instead, we compute the search direction $-H_k\nabla f_k$ directly by a \emph{L-BFGS two-loop recursion} algorithm (Algorithm~\ref{alg:twoloop})~\cite{Nocedal1999}:
\begin{algorithm}[h]
\caption{L-BFGS two-loop recursion}
\label{alg:twoloop}
\begin{algorithmic}
\STATE $q = \nabla f(x_k)$;
\FOR {$i=k-1, k-2,\ldots, k-m$}
\STATE $\alpha_i = \rho_i s^T_i q$;
\STATE $q = q - \alpha_i y_i$;
\ENDFOR
\STATE $z = H^{0}_k q$;
\FOR {$i=k-m, k-m+1, \ldots k-1$}
\STATE $\beta_i = \rho_i y^T_i z$;
\STATE $z = z + s_i (\alpha_i - \beta_i)$;
\ENDFOR
\STATE Output $z$ to be $H_k\nabla f_k$.
\end{algorithmic}
\end{algorithm}

The L-BFGS optimization procedure is described in Algorithm~\ref{lbfgsalgorithm}~\cite{Nocedal1999}.
\begin{algorithm}[h]
\caption{the L-BFGS algorithm}
\label{alg:LBFGS}
\begin{algorithmic}
\STATE Choose a starting point $x_0$ and a positive integer $m$;
\STATE $k$ = 0;
\REPEAT
\STATE Choose $H_k^0$;
\STATE Compute a descending direction $p_k$ by the two-loop recursion algorithm;
\STATE Compute $x_{k+1}=x_k+\alpha_k p_k$, $\alpha_k$ is chosen to satisfy the Wolfe conditions;
\IF{$k>m$}
\STATE Discard $s_{k-m}$ and $y_{k-m}$;
\ENDIF
\STATE Compute the values of $s_k$, $y_k$ and store them;
\UNTIL convergence
\end{algorithmic}
\label{lbfgsalgorithm}
\end{algorithm}

L-BFGS stops when the norm of the gradient of the objective function is smaller than a specified tolerance value $\epsilon$, i.e. $\|\nabla f\|<\epsilon$.

In algorithm~\ref{lbfgsalgorithm}, once the descending direction $p_k$ is obtained, the variables $x_{k+1}$ should be updated by $x_{k+1} = x_{k} + \alpha_k p_{k}$.  Here $\alpha_k$ is chosen to guarantee the decreasing of the value of the objective function. This is usually solved by a linesearch algorithm. Basically, a linesearch algorithm starts with $\alpha_k = 1$ and decreases the value of $\alpha_k$ by some strategy until the following \emph{Wolfe conditions} are satisfied~\cite{Nocedal1999}:
\[
\left\{
\begin{array}{l}
f(x_k+\alpha_k p_k)\leq f(x_k)+c_1\alpha_k\nabla f^T_kp_k,\\
\nabla f(x_k+\alpha_kp_k)^Tp_k\ge c_2\nabla f^T_kp_k.
\end{array}
\right.
\]
Here $c_1$ and $c_2$ are constants which satisfy $0<c_1<c_2<1$. In our algorithm we use $c_1=10^{-4}$ and $c_2=0.9$ throughout optimization. We have found from our experience that $\alpha_k =1$ is often good enough and the computational time of linesearch only takes a small partition of the total computational time. We will show these timing data in later sections.

\FloatBarrier

\section{Curve fitting with L-BFGS}

\subsection{Algorithm outline}
We employ L-BFGS directly to solve the nonlinear least squares minimization problem~\ref{eq:general} in the following steps.

\begin{enumerate}
\item  Specify an initial curve $\mathbf{P}(t)$.
\item  Find the foot point $\mathbf{P}(t_k)$ on $\mathbf{P}(t)$ for every data point $\mathbf{X}_k$. This gives the initial position of location parameters $\{t_k\}$.
\item  Run Algorithm~\ref{lbfgsalgorithm}: the L-BFGS algorithm until convergence.
\end{enumerate}

We will refer to this algorithm as the \emph{L-BFGS fitting method} in the following sections.

\subsection{Selection of $m$}
%

The L-BFGS algorithm (Algorithm~\ref{lbfgsalgorithm}) uses $m$ gradient vectors in a sequence of iterations to approximate the inverse Hessian matrix. A larger $m$ can result in a more accurate approximation  but at the same time, it will take more computational time.  Therefore, it is important to select a proper value of $m$ which balances between the competing objectives, i.e., accuracy and efficiency. In literature, the value of $m$ is often chosen between 3 and 20~\cite{Nocedal1999}. However there are also papers reporting that a very large value of $m$ is necessary for generating satisfactory results. One example is~\cite{jiang2004preconditioned} in which $m$ is set to 240.

To find a proper value of $m$ in our curve fitting problem, we have tested many examples with $m=3$, $5$, $7$, $20$, $50$ and $120$ to understand the behavior of our algorithm. Our conclusion is, for simple examples, using $m=3$ and $m=5$ may lead to a slightly faster fitting error descending speed. On the other hand, using $m=20$, $50$ and $120$ will contribute to a faster gradient norm convergence speed.
For more complicated examples (examples with more data points and heavier noise), the behaviors of error descending speed and gradient norm descending speed using  different $m$ tend to be indistinguishable. Considering all these factors, we suggest using $m=20$ in experiments.

\subsection{Foot point projection}

Computing foot points needs to be performed in every iteration of traditional curve fitting methods. Foot point projection is itself an optimization problem which is investigated in~\cite{Hoschek:1989:OAC:80732.80734},~\cite{Saux:2003:IHI:965907.965910},~\cite{jue:05e} and~\cite{Hu2005}:
\[
 \min_{t}\| \mathbf{P}(t) - \mathbf{X}_k \|^2.
\]
When implementing the traditional methods we compute foot points in every iteration using the Gauss-Newton method outlined below.

1. \emph{Initialization}: In this step location parameters corresponding to data points are roughly estimated. The estimation will be used as initialization for further optimization in the next step. A straightforward method is: Sampling dense enough points on the fitting curve and finding the closest sample to each data point as the estimation of foot points.

2. \emph{Iterative Update}: Update the location parameter iteratively~\cite{Wang2006}: in the $i$-th iteration, the location parameter of $\mathbf{X}_k$ is updated by  $t_{k,i+1} =  t_{k,i} + a {\delta}t$, where
\[{\delta}t =\frac{(\mathbf{X}_k-\mathbf{P}(t_{k,i}))\cdot\mathbf{P}'(t_{k,i})}{\|\mathbf{P}'(t_{k,i})\|^2}\]
is the suggested update in the descending direction. The value of $a$ is decided by a simple linesearch method to guarantee the decreasing of the orthogonal distance, i.e., $\|\mathbf{P}(t_{k,i+1})-\mathbf{X}_k\|<\|\mathbf{P}(t_{k,i})-\mathbf{X}_k\|$. The optimization process stops when  $\|(\mathbf{X}_k-\mathbf{P}(t_{k,i})) \cdot \mathbf{P}'(t_{k,i})\| < 10^{-10}$.

The \textit{Initialization} step is generally time consuming. In practice, after several iterations in the beginning of curve fitting, the shape of the fitting curve will not change a lot in later optimization.  In this case, we can use foot points computed in the $i$-th iteration  (i.e.,$\{t_{k,i}\}$) as initialization foot points for the $(i+1)$-th iteration. To determine whether it is safe to directly use the foot points from the last iteration as initialization, we use a criterion based on variation of fitting error: We suggest to re-initialize foot points in the $(i+1)$-th iteration when $\frac{|E_{i+1}-E_{i}|}{E_{i+1}}>0.2$ where $E_{i}$ and $E_{i+1}$ are the fitting error in the  $i$-th iteration and $(i+1)$-th iteration respectively, as defined in Eqn.~\ref{eq:error_def}. In our experiments, with this criterion, for all traditional methods, the number of foot point initializations needed is from 1 to 4 in all our tested examples.

\subsection{Foot point correction and restart of L-BFGS}

\begin{figure}[htb]
\begin{minipage}{0.45\textwidth}
\vspace{-0.2cm}
\centering
\subfigure[Without foot point correction.]{\includegraphics[width=0.3\textwidth]{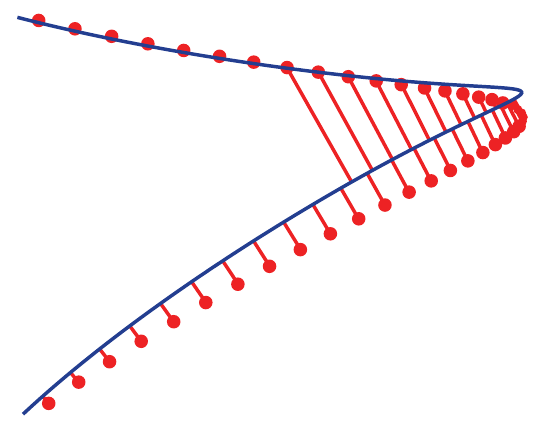}}
\hspace{1cm}
\subfigure[With foot point correction.]{\includegraphics[width=0.3\textwidth]{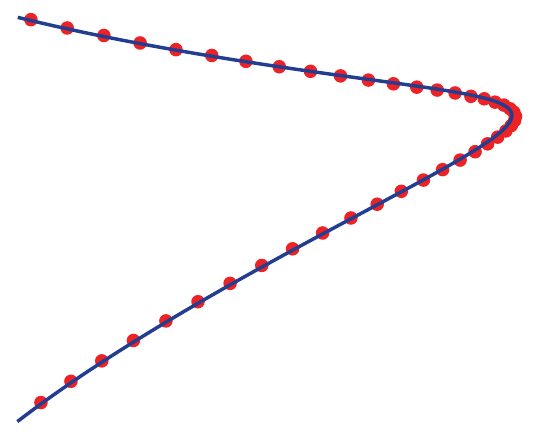}}
\caption{foot point correction.}
\label{fig:restart}
\vspace{-0.5cm}
\end{minipage}
\end{figure}

The initialization of the L-BFGS fitting method also needs foot point projection to determine the initial set $\mathcal{T}$ of location parameters. Then, in the subsequent iterations the L-BFGS fitting method in general does not need to perform foot point projection any more but rather optimizes location parameters $\{ t_k \}$ and control points $\mathcal{P}$ simultaneously. In rare case, especially when the initial fitting curve specified is not good enough, it is possible that $\mathbf{P}(t_k)$  is far from the closet point on the curve to $\mathbf{X}_k$, even if $\mathbf{P}(t_k) - \mathbf{X}_k$ is orthogonal to $\mathbf{P}'(t_k)$. An example is shown in Figure~\ref{fig:restart}(a): it is part of a fitting curve on convergence, but there are points $\mathbf{P}(t_k)$ which are not the foot points of $\mathbf{X}_k$, because the L-BFGS fitting method gets stuck in a poor local minimum. In this case the following remedy can be used. 

To rectify the incorrect foot point projections, we just perform foot point computation after the termination of current L-BFGS algorithm, and start a new L-BFGS algorithm taking initial control points from the previous run of L-BFGS and initial location parameters from the output of foot point computation. Figure~\ref{fig:restart}(b) shows the result fitting curve. To detect a local minimum automatically, we measure the fitting error $E$ after the L-BFGS algorithm and the fitting error $E_{+}$ after the additional operation of foot point computation. If the error changing $\| E - E_{+}\|$ is bigger than a tolerance (we use $10^{-6}$ in our experiments), we conclude that the foot points of some data points are corrected and an additional run of the L-BFGS algorithm is needed.

\section{Results and Discussions}
\begin{figure}[t!]
\begin{minipage}{0.45\textwidth}
\centering
\subfigure[Initialization.]{\includegraphics[height=3cm]{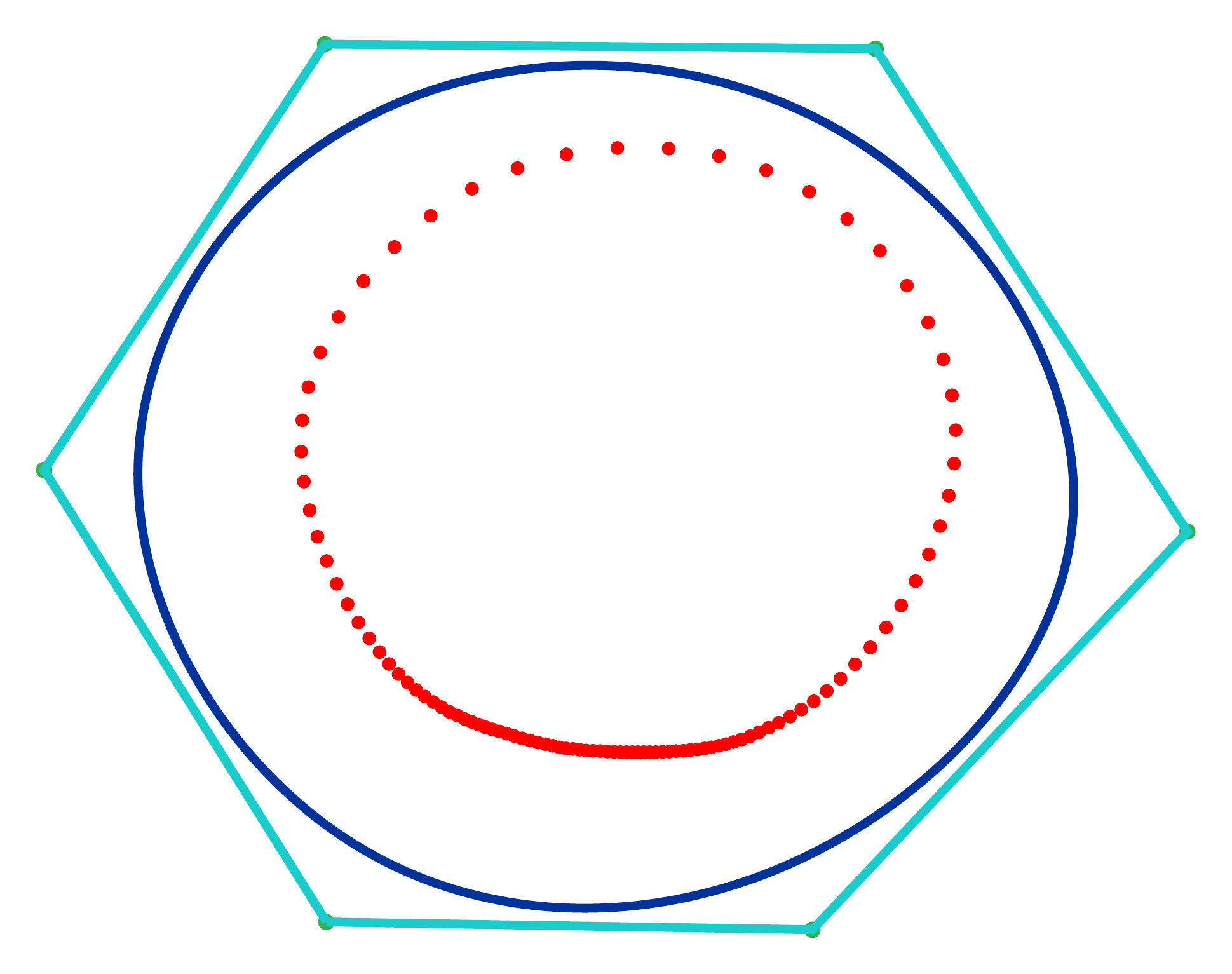}}
\subfigure[Fitting curve.]{\includegraphics[height=3cm]{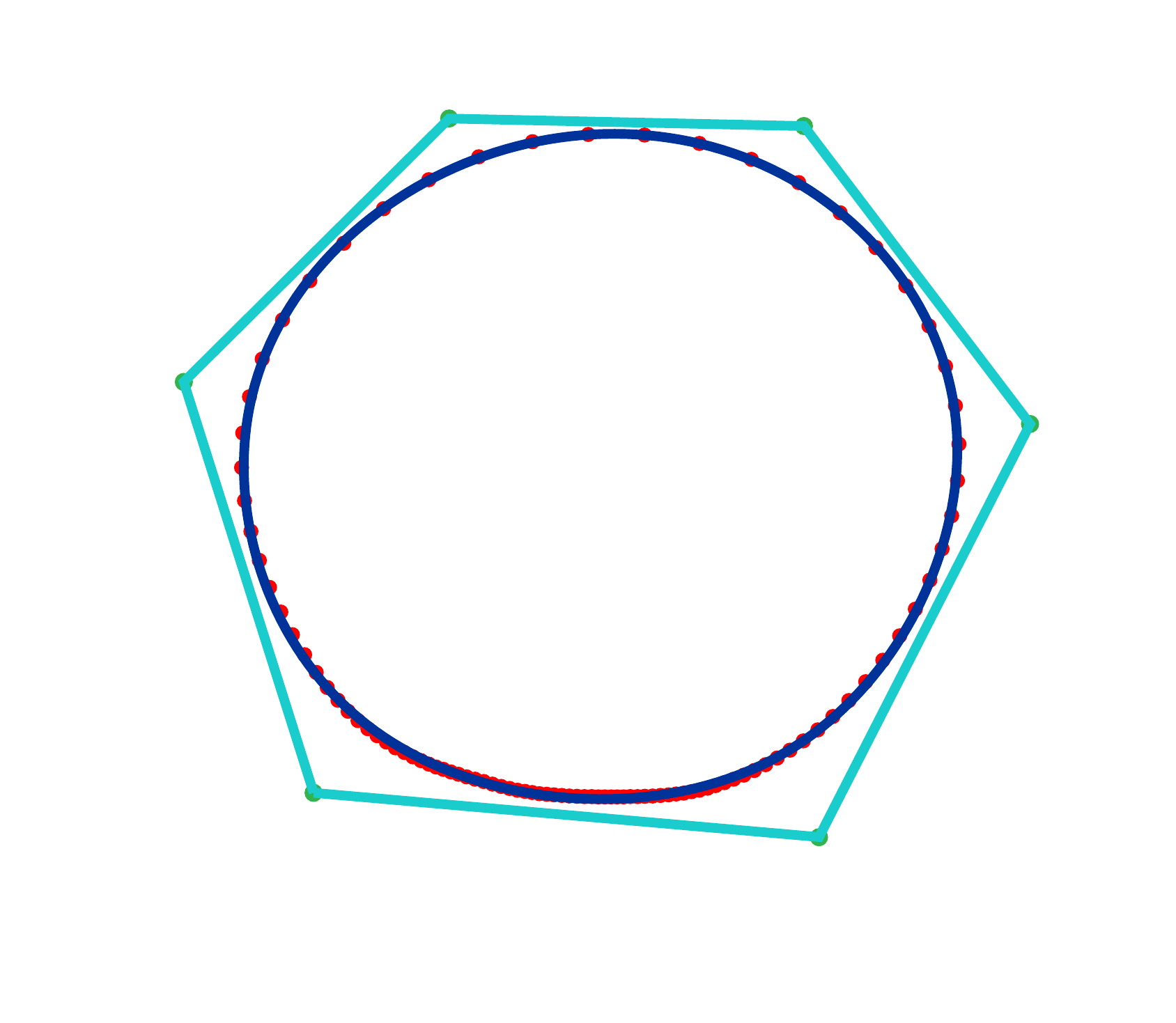}}\\
\vspace{-0.3cm}
\subfigure[Error vs iteration.]{\includegraphics[width=0.8\textwidth]{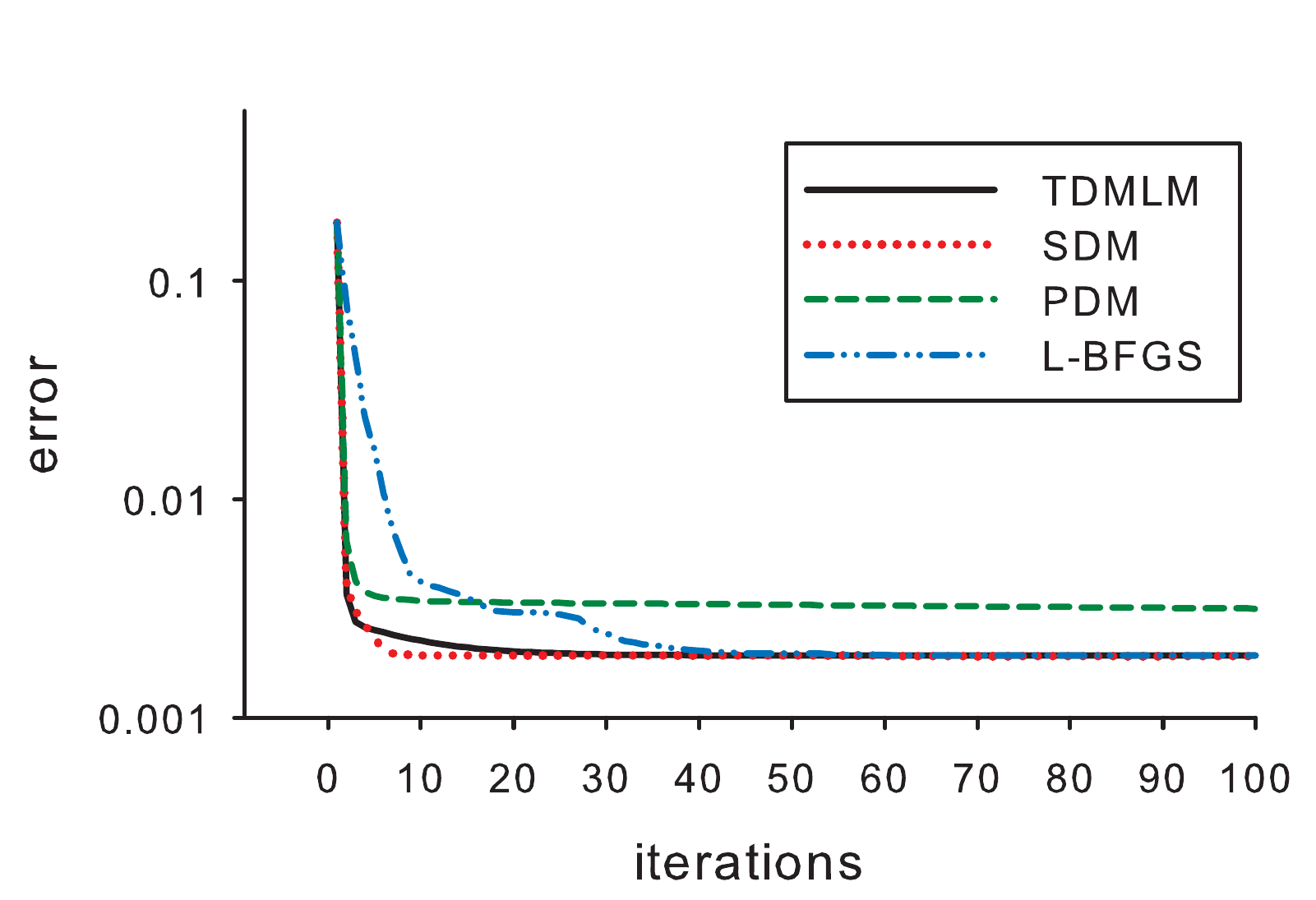}}\\
\vspace{-0.4cm}
\subfigure[Error vs time.]{\includegraphics[width=0.8\textwidth]{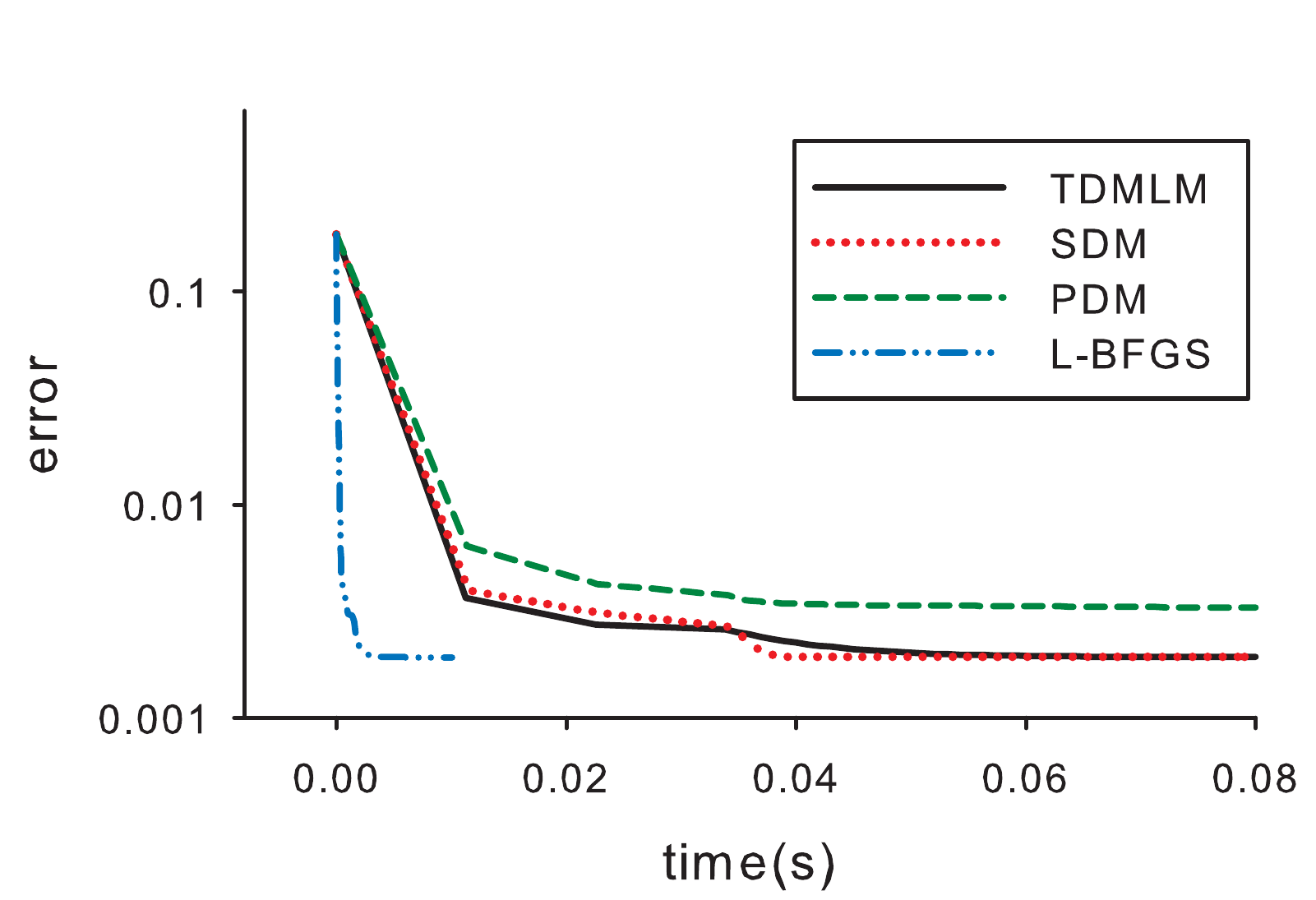}}\\
\vspace{-0.4cm}
\subfigure[Gradient norm vs time.]{\includegraphics[width=0.8\textwidth]{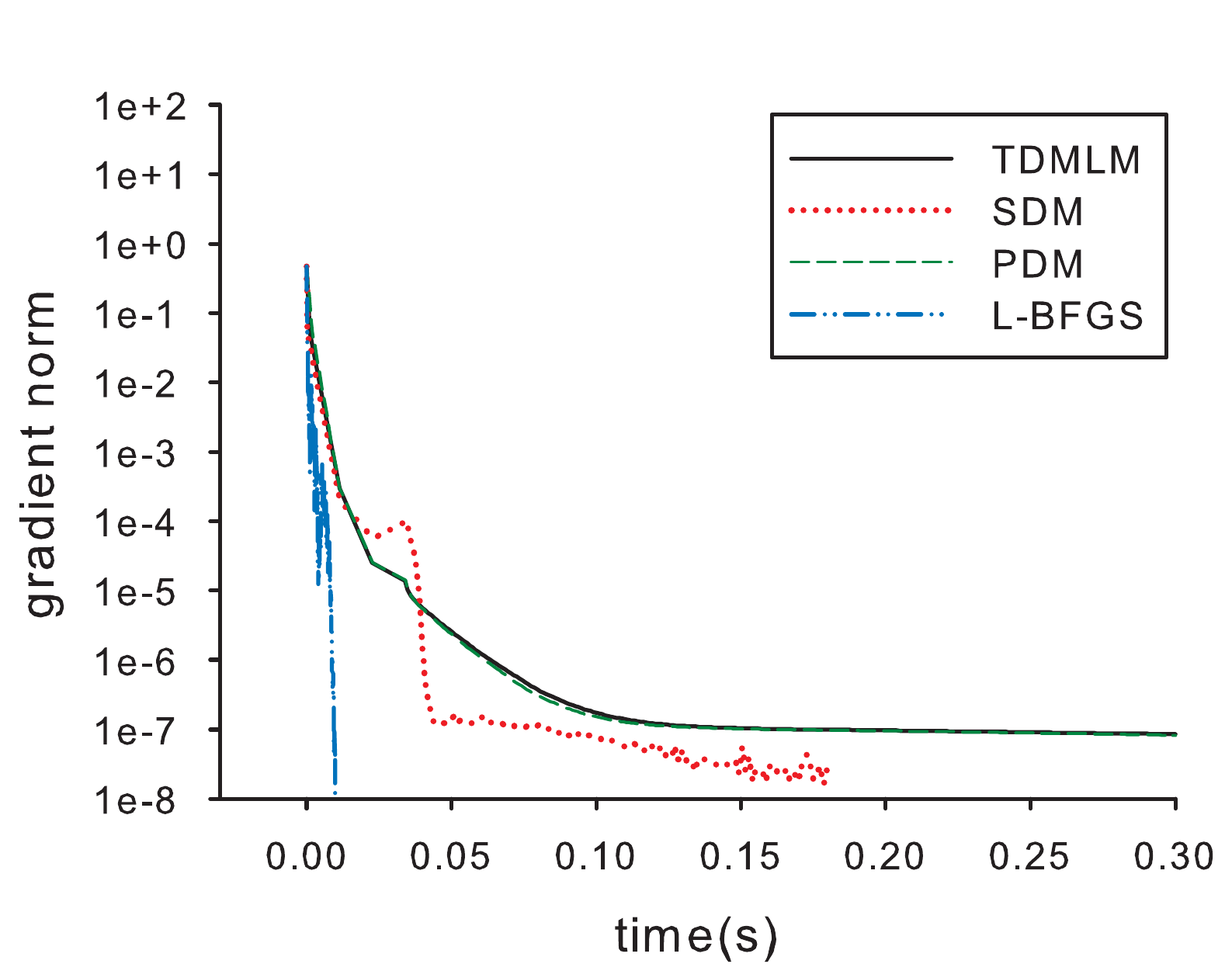}}\\
\vspace{-0.3cm}
\subfigure[Time to attain minimal error (in seconds).]{
\begin{tabular}{cccc}
\hline
L-BFGS & PDM & TDMLM & SDM \\
$3.2\cdot 10^{-3}$ & $0.95$ & $4.5\cdot 10^{-2}$ & $4.0\cdot 10^{-2}$\\
\hline
\end{tabular}
}\\
\subfigure[Time cost for an iteration (in seconds).]{
\begin{tabular}{cccc}
\hline
L-BFGS & PDM & TDMLM & SDM \\
$5.94\cdot10^{-5}$ & $9.61\cdot 10^{-4}$ & $1.04\cdot 10^{-3}$ & $1.41\cdot 10^{-3}$\\
\hline
\end{tabular}}
\caption{The target shape is a set of 100 points on a circle. A B-spline curve with 6 control points is used to fit it. No fairing term is used in this example. }
\label{fig:s9}
\vspace{-0.4cm}
\end{minipage}
\end{figure}
\begin{figure}[t!]
\begin{minipage}{0.45\textwidth}
\centering
\subfigure[Initialization.]{\includegraphics[height=3cm]{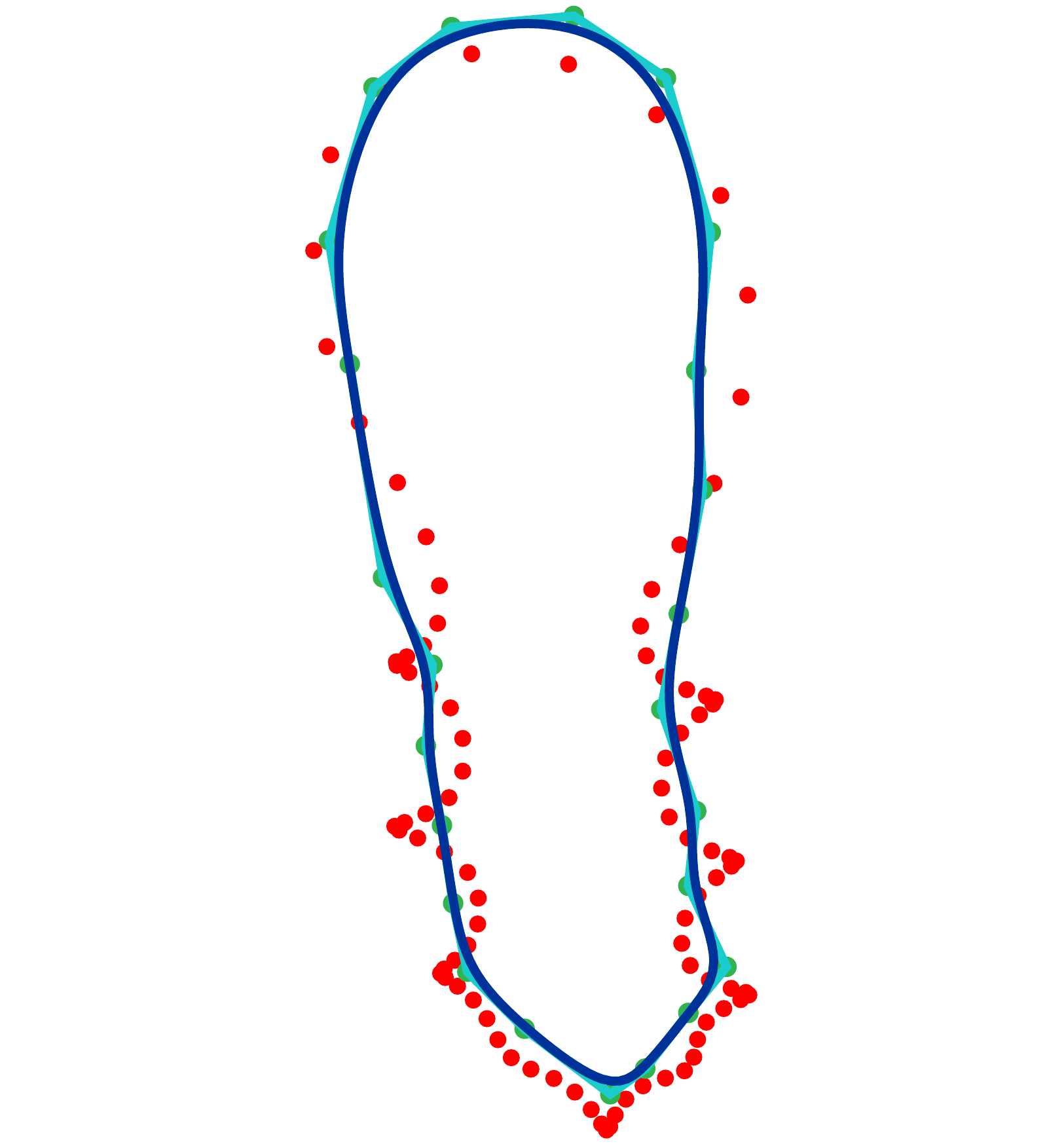}}
\subfigure[Fitting curve.]{\includegraphics[height=3cm]{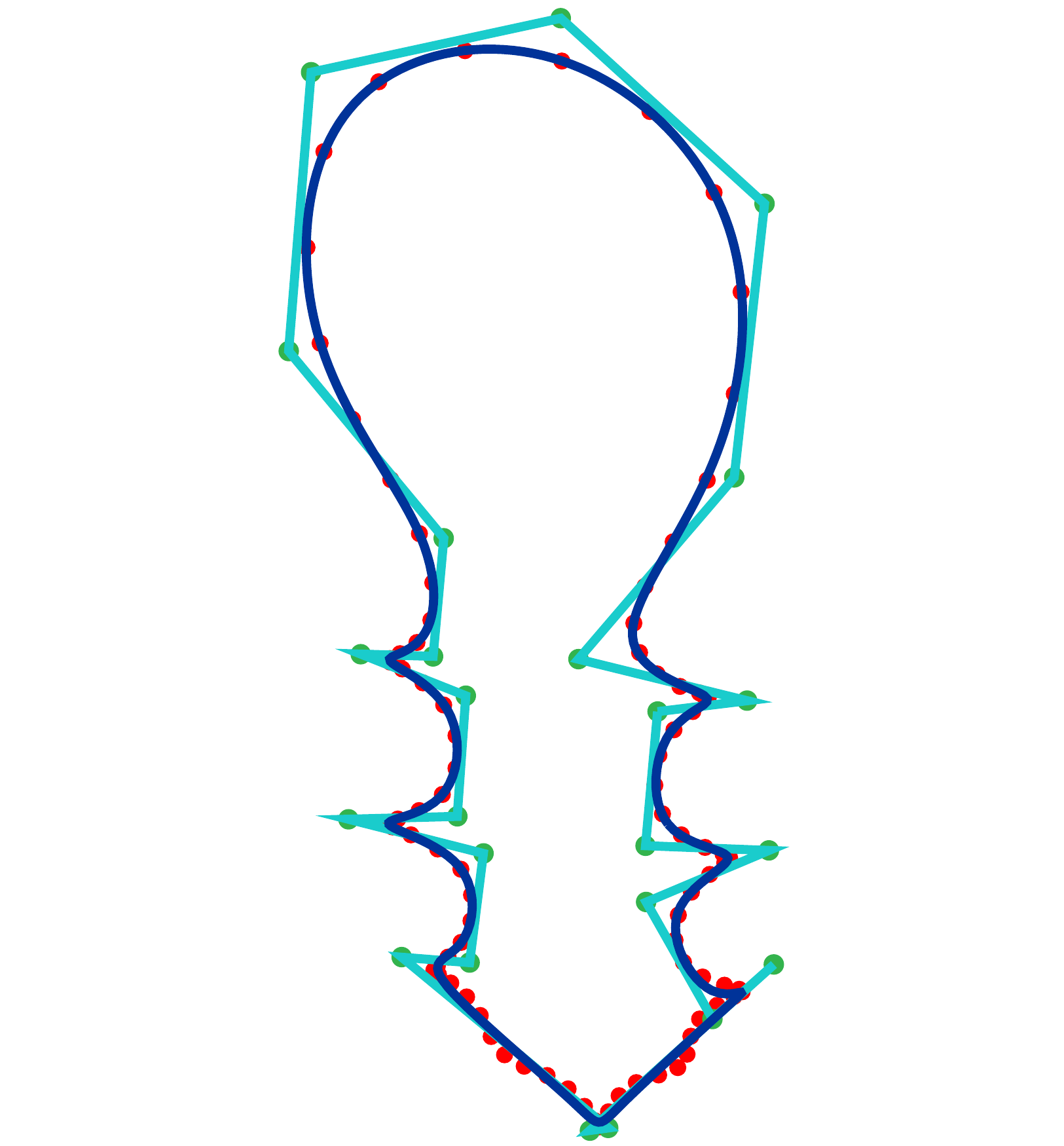}}\\
\vspace{-0.3cm}
\subfigure[Error vs iteration.]{\includegraphics[width=0.8\textwidth]{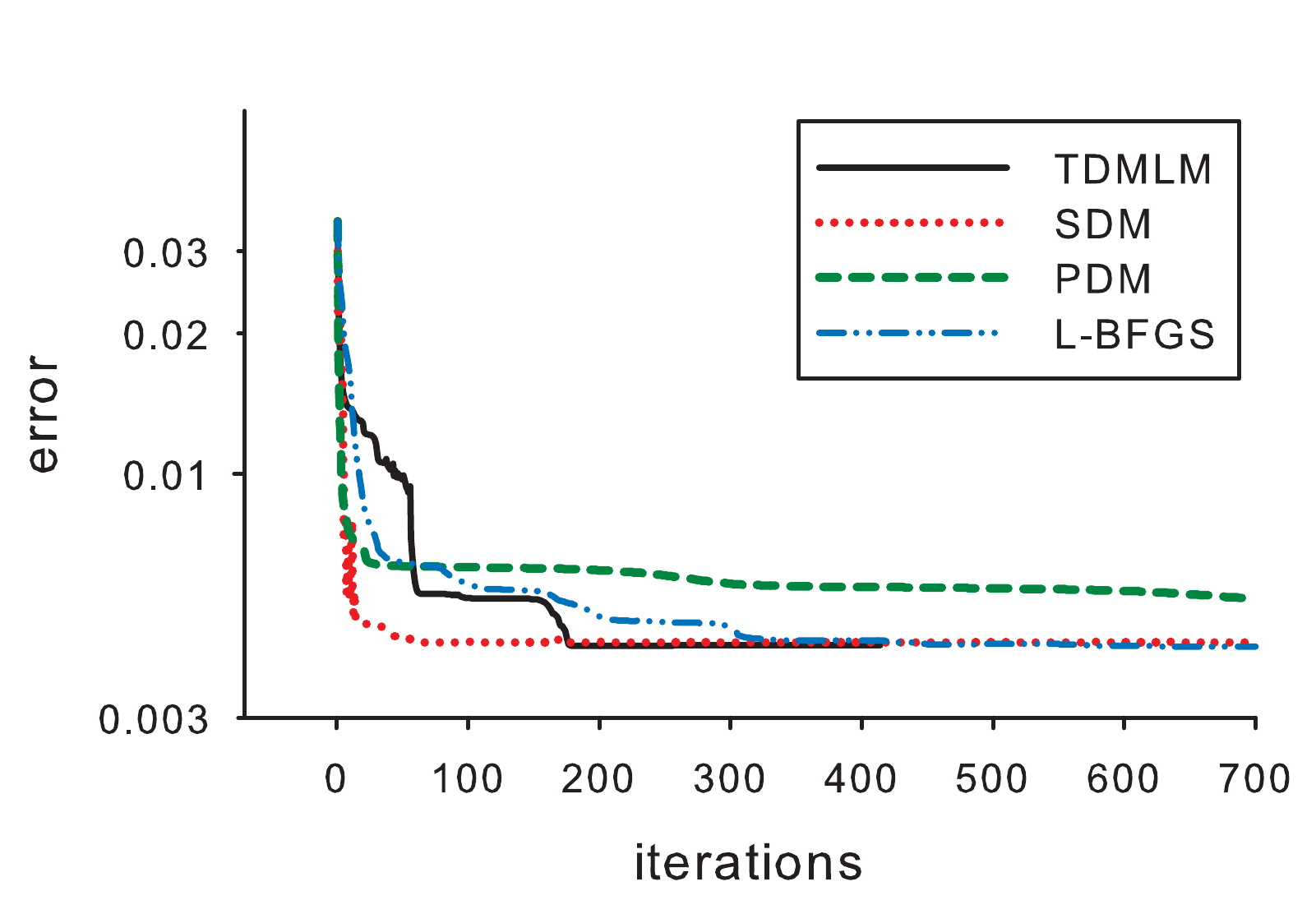}}\\
\vspace{-0.4cm}
\subfigure[Error vs time.]{\includegraphics[width=0.8\textwidth]{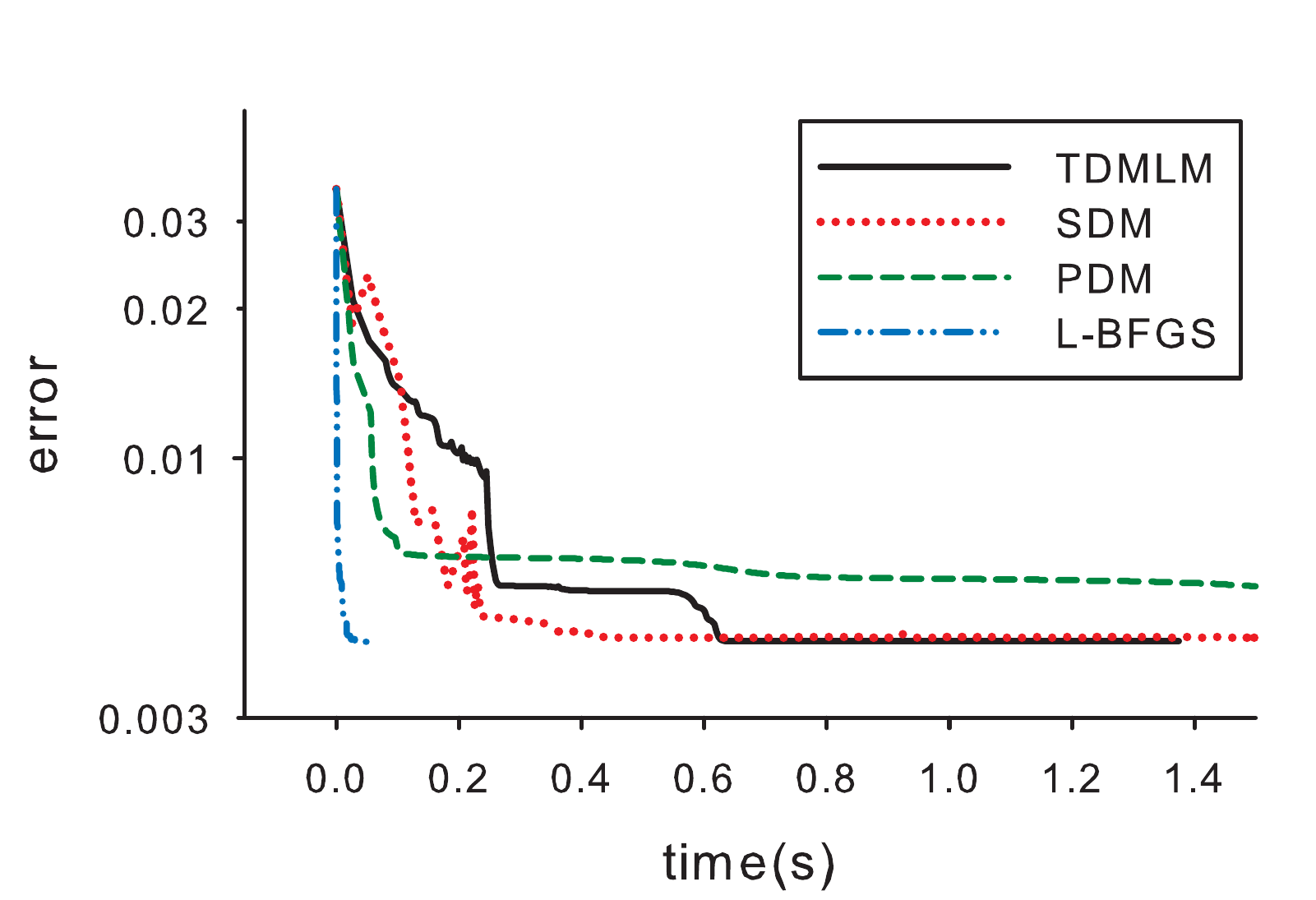}}\\
\vspace{-0.4cm}
\subfigure[Gradient norm vs time.]{\includegraphics[width=0.8\textwidth]{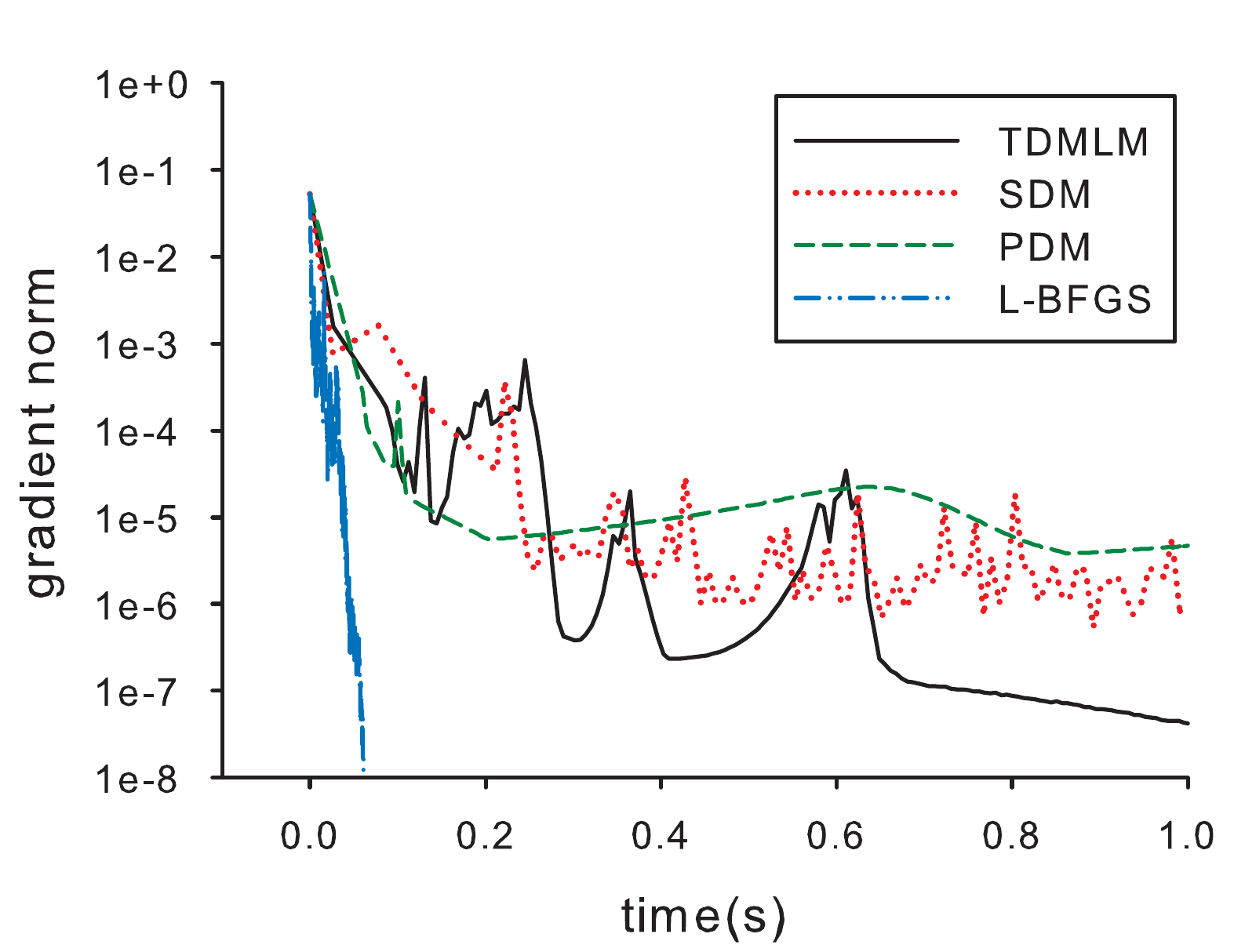}}\\
\vspace{-0.3cm}
\subfigure[Time to attain minimal error (in seconds).]
{
\begin{tabular}{cccc}
\hline
L-BFGS & PDM & TDMLM & SDM \\
$3.5\cdot 10^{-3}$ & $3.3$ & $0.63$ & $0.43$\\
\hline
\end{tabular}
}\\
\subfigure[Time cost for an iteration (in seconds).]{
\begin{tabular}{cccc}
\hline
L-BFGS & PDM & TDMLM & SDM \\
$5.39\cdot 10^{-5}$ & $2.22\cdot 10^{-3}$ & $3.32\cdot 10^{-3}$ & $4.64\cdot 10^{-3}$\\
\hline
\end{tabular}}
\caption{An example with sharp features. The fitting curve has 24 control points and the data set contains 90 points. The coefficients of fairing term are set to $\alpha=0$ and $\beta=5\cdot 10^{-4}$. }
\label{fig:n4}
\vspace{-0.3cm}
\end{minipage}
\end{figure}
\begin{figure}[t!]
\begin{minipage}{0.45\textwidth}
\centering
\subfigure[Initialization.]{\includegraphics[height=3cm]{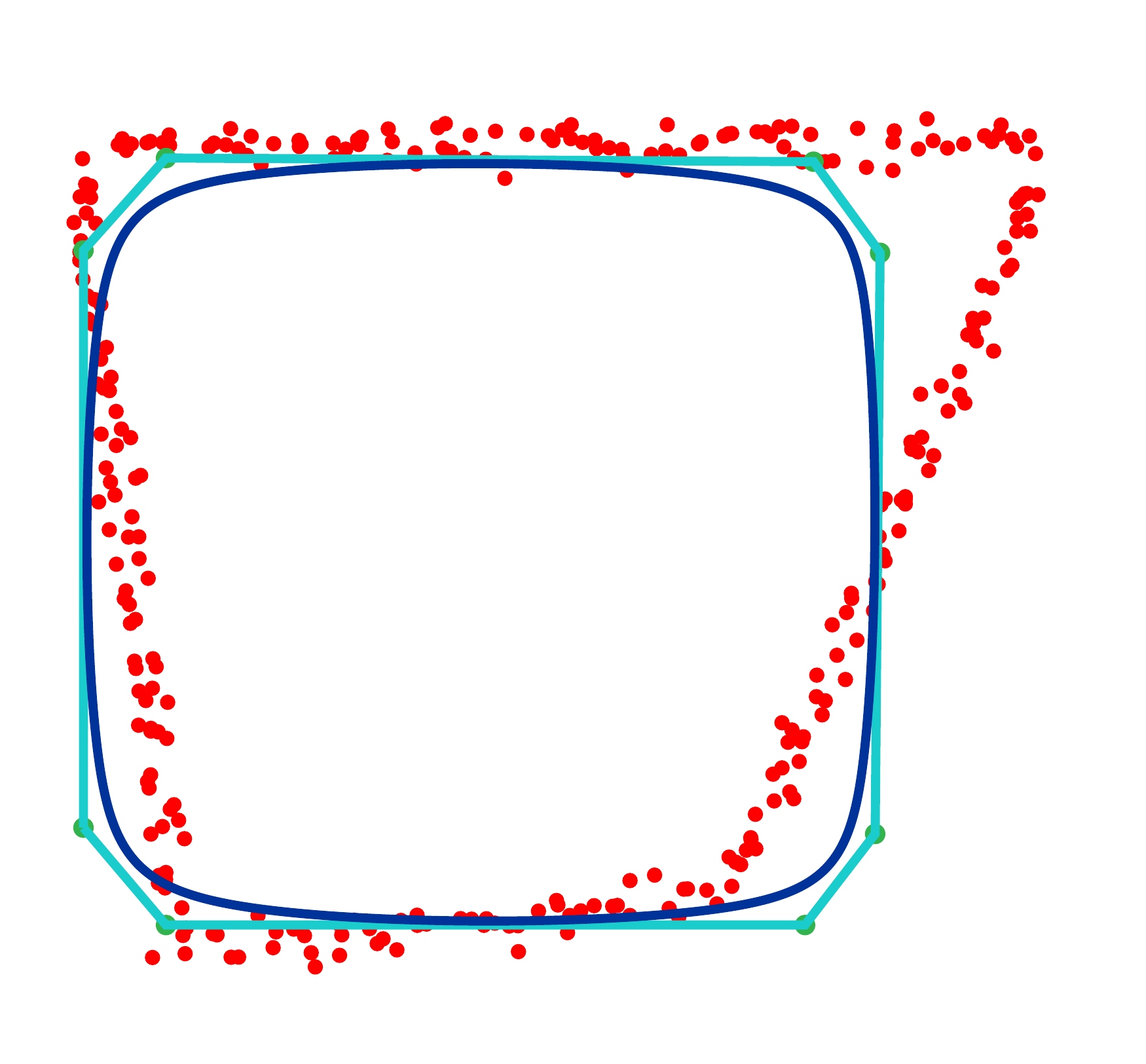}}
\subfigure[Fitting curve.]{\includegraphics[height=3cm]{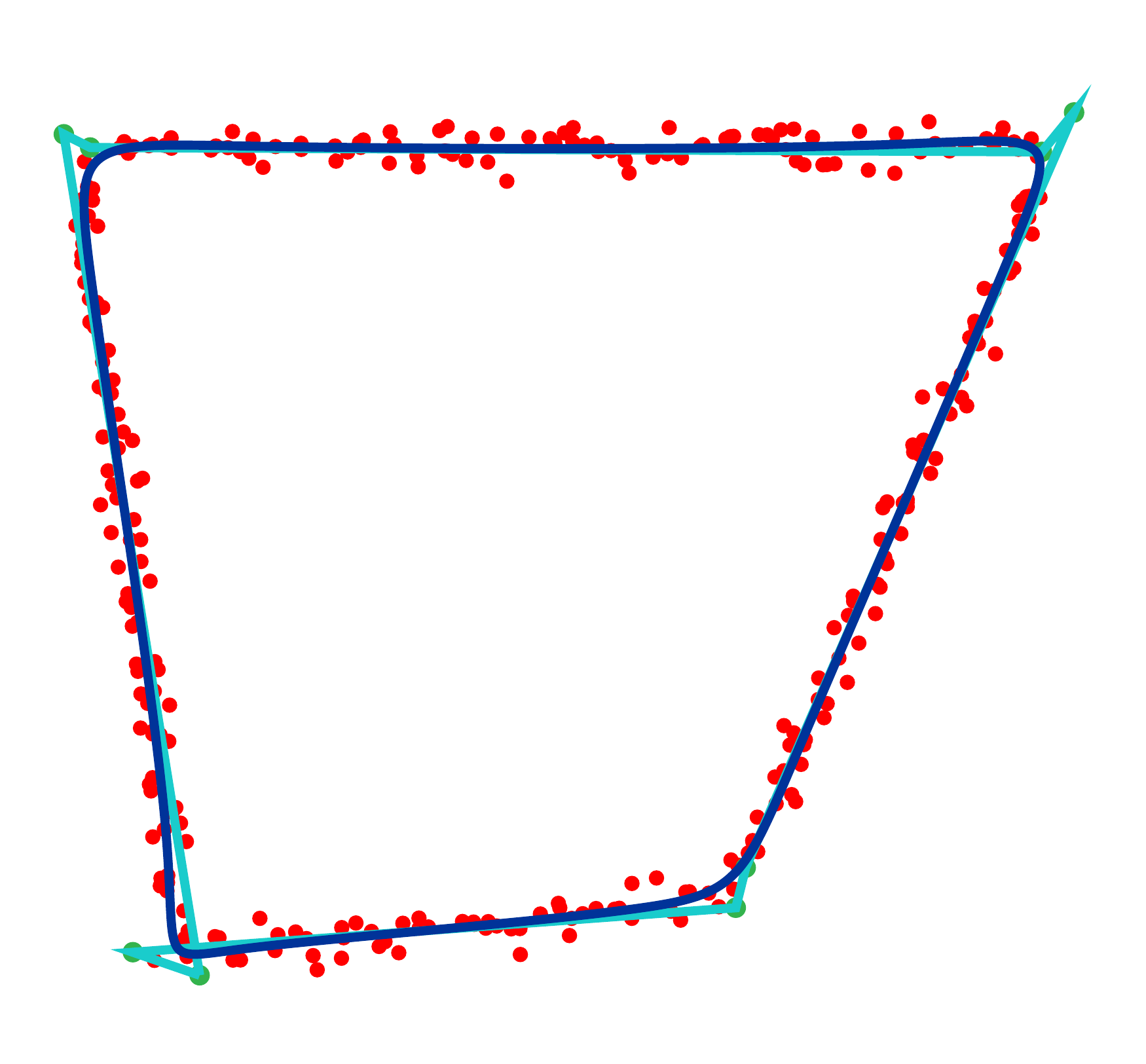}}\\
\vspace{-0.4cm}
\subfigure[Error vs iteration.]{\includegraphics[width=0.8\textwidth]{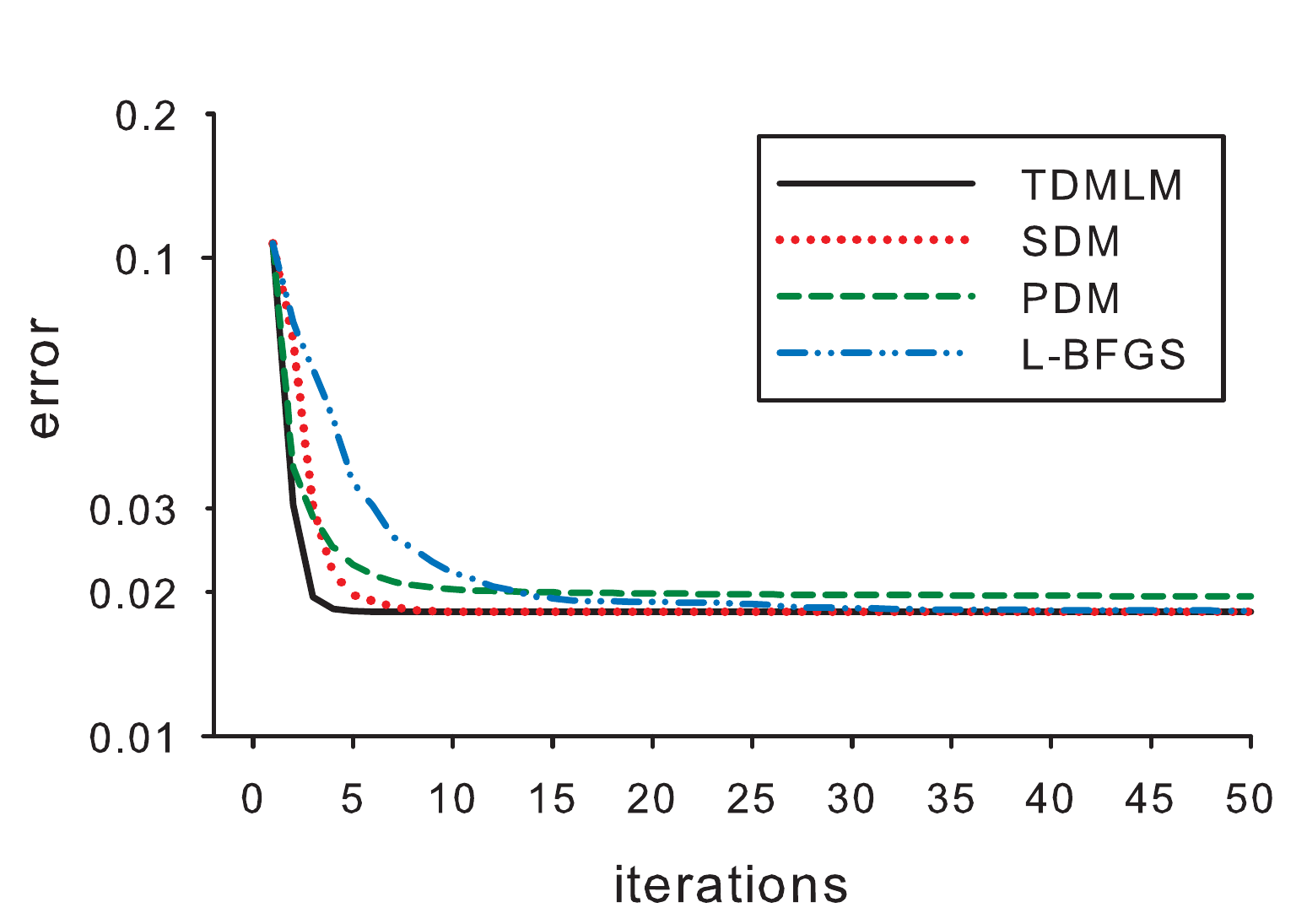}}\\
\vspace{-0.4cm}
\subfigure[Error vs time.]{\includegraphics[width=0.8\textwidth]{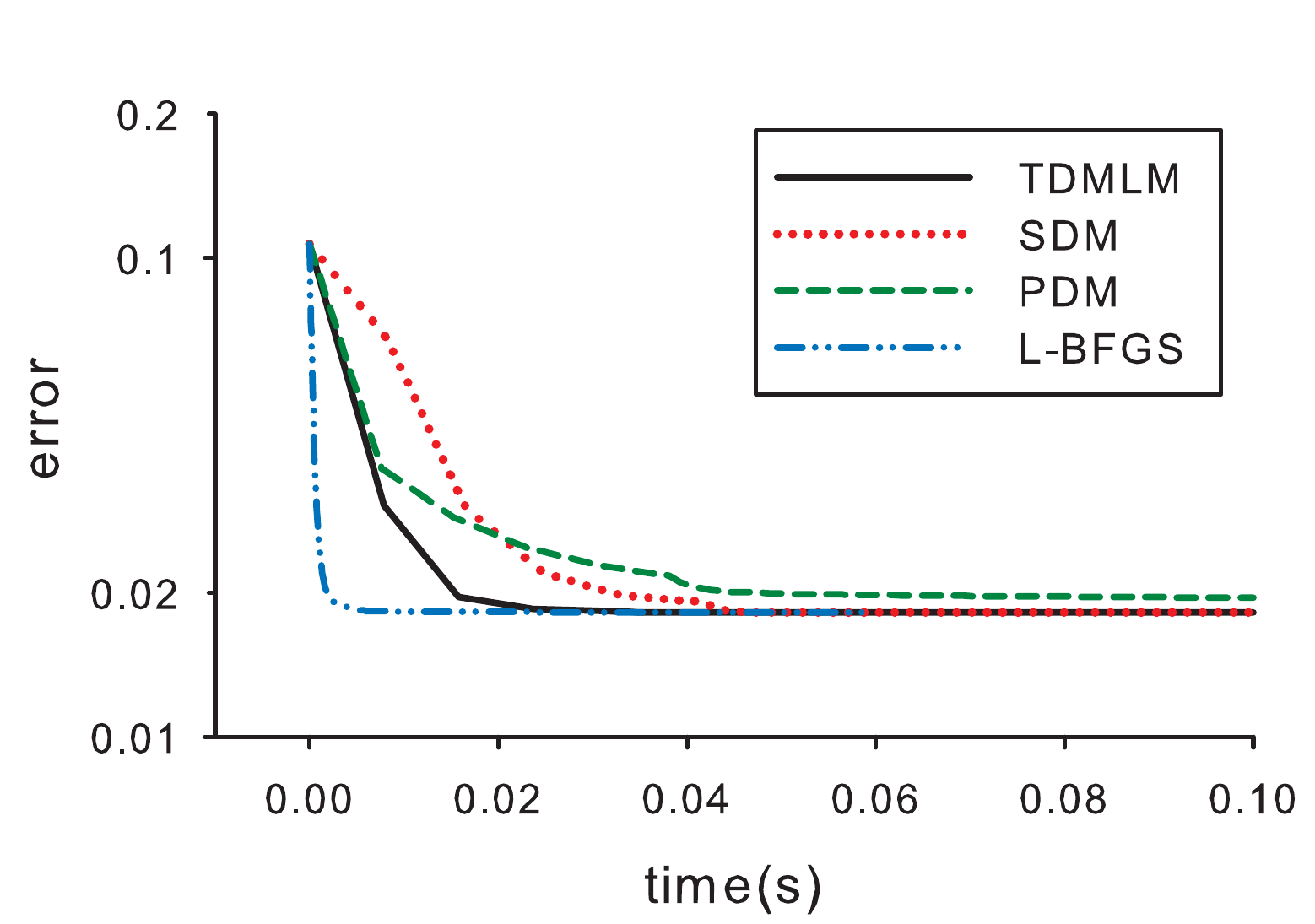}}\\
\vspace{-0.4cm}
\subfigure[Gradient norm vs time.]{\includegraphics[width=0.8\textwidth]{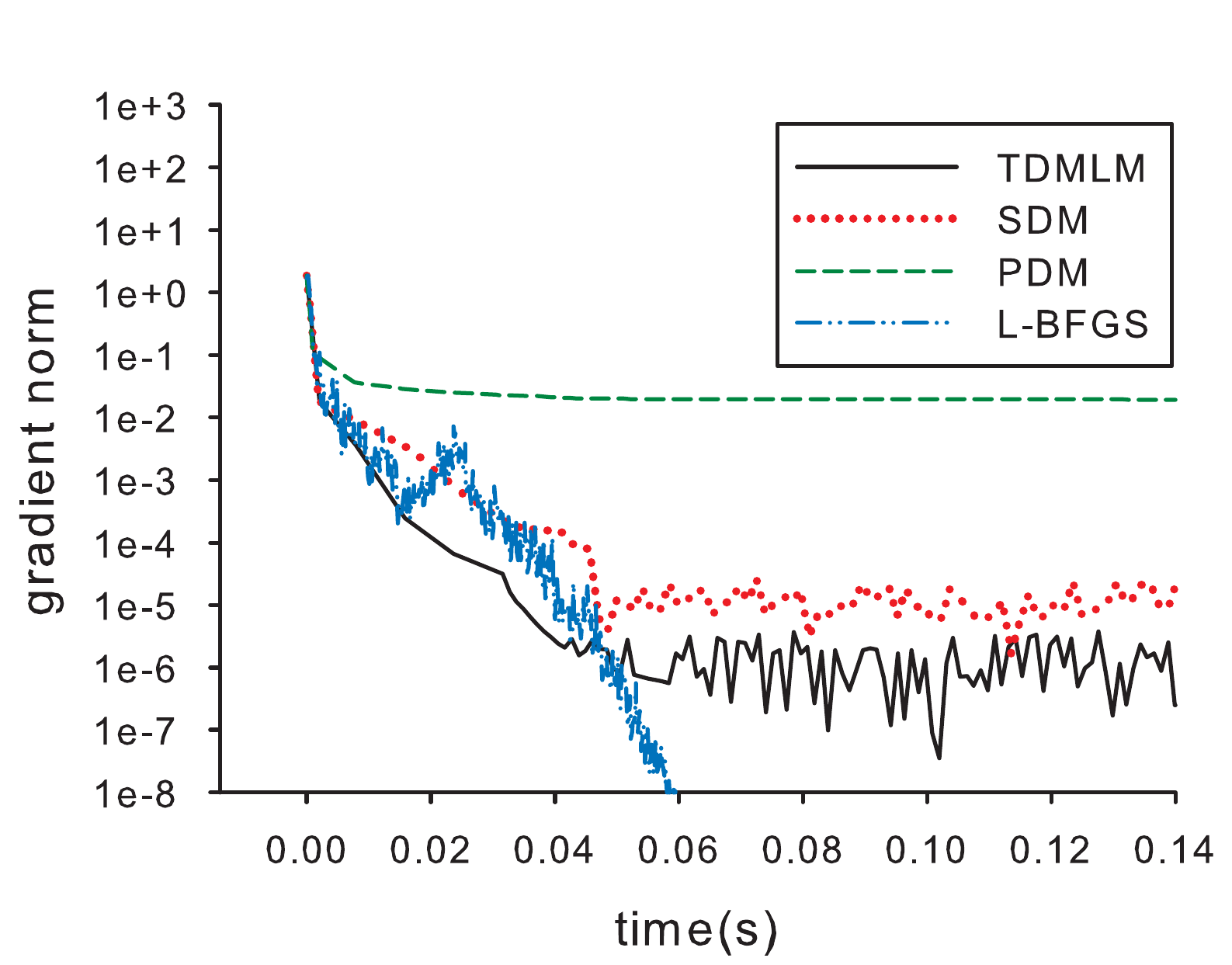}}\\
\vspace{-0.2cm}
\subfigure[Time to attain minimal error (in seconds).]{
\begin{tabular}{cccc}
\hline
L-BFGS & PDM & TDMLM & SDM \\
$5.2\cdot 10^{-3}$ &$ 0.82$ & $3.2\cdot 10^{-2}$ & $4.4\cdot 10^{-2}$\\
\hline
\end{tabular}}\\
\subfigure[Time cost for an iteration (in seconds).]{
\begin{tabular}{cccc}
\hline
L-BFGS & PDM & TDMLM & SDM \\
$1.52\cdot 10^{-4}$ & $9.07\cdot10^{-4}$ & $1.11\cdot 10^{-3}$ & $1.14\cdot 10^{-3}$\\
\hline
\end{tabular}
}
\caption{An example with noisy data points. The fitting curve has 8 control points and the data set has 150 points. No fairing term is used in this example. }
\vspace{-0.3cm}
\label{fig:9}
\end{minipage}
\end{figure}

In this section, we first present some experiments comparing the L-BFGS fitting method with existing methods, then we give explanation on the fast speed of the L-BFGS fitting method.

\subsection{Experiments}

The fitting error of the $i$-th iteration is measured by:
\begin{equation}
\label{eq:error_def}
E_i:=\left(\sum\limits_{k=1}^N \frac{1}{N}\|\mathbf{P}(t_k)-\mathbf{X}_k\|^2\right)^\frac{1}{2}.
\end{equation}

The parameter domains of B-spline curves in these examples are set to $[0,1]$. All data points are scaled into a unit box: $[0,1]\times[0,1]$. 

Due to different complexities and the set up of initial curves in the examples in our experiments, we use different coefficients of fairing terms $\alpha$ and $\beta$ in different examples to obtain satisfactory fitting curves.  In each example, the same values of faring term coefficients  are used for all tested methods. The values of coefficients are noted in the captions of Figures. 

\textbf{Comparison with traditional methods.} Three data sets are given in Figure~\ref{fig:s9},~\ref{fig:n4} and~\ref{fig:9} for comparisons with three traditional methods: PDM, TDMLM and SDM. For each data set, we show data points, the initial fitting curve and the final fitting curve of the L-BFGS fitting method. Three charts are also shown for each data set. The first two charts show the fitting error versus the iteration number and computational time respectively. The third chart shows the decreasing of gradient norm versus computational time.

We observe that in the first several iterations, the fitting error of the L-BFGS fitting method does not decrease as fast as SDM and TDMLM in terms of number of iterations. That is because that in the L-BFGS algorithm (Algorithm~\ref{lbfgsalgorithm}), the approximation of inverse Hessian matrix needs to be accumulated by using information from a sequence of $m$ iterations. Therefore, at the first several iterations, the approximant matrix is not accurate enough and this slows down the performance of the   L-BFGS fitting method. However, an iteration of the L-BFGS fitting method is much faster than PDM, TDMLM and SDM. As a result, the L-BFGS fitting method converges much faster than the other three methods in terms of computational time, as shown in
Figure~\ref{fig:s9}(d),~\ref{fig:n4}(d) and~\ref{fig:9}(d).

\textbf{Convergence.} The convergence behaviors of the four methods can be observed from the third chart in the above three examples, showing the decrease of gradient norm against computational time. The termination criterion for all these examples is $\|\nabla f\|_\infty< 10^{-8}$, where $f$ is the objective function.  From Figure
~\ref{fig:s9}(e),~\ref{fig:n4}(e) and~\ref{fig:9}(e), we observe that the L-BFGS fitting method is the only method that always meets this criterion, i.e. the gradient norm of its objective function reaches the threshold of $10^{-8}$. This is not surprising since the L-BFGS fitting method implements a well-studied optimization method (the L-BFGS algorithm, Algorithm~\ref{lbfgsalgorithm}) that has demonstrated superior convergence behavior close to the superlinear rate possessed by the BFGS method~\cite{Nocedal1999}.
\begin{figure}[htbp]
\begin{minipage}{.49\textwidth}
\vspace{-0.2cm}
\centering
\subfigure[Initialization.]{\includegraphics[height=2.5cm]{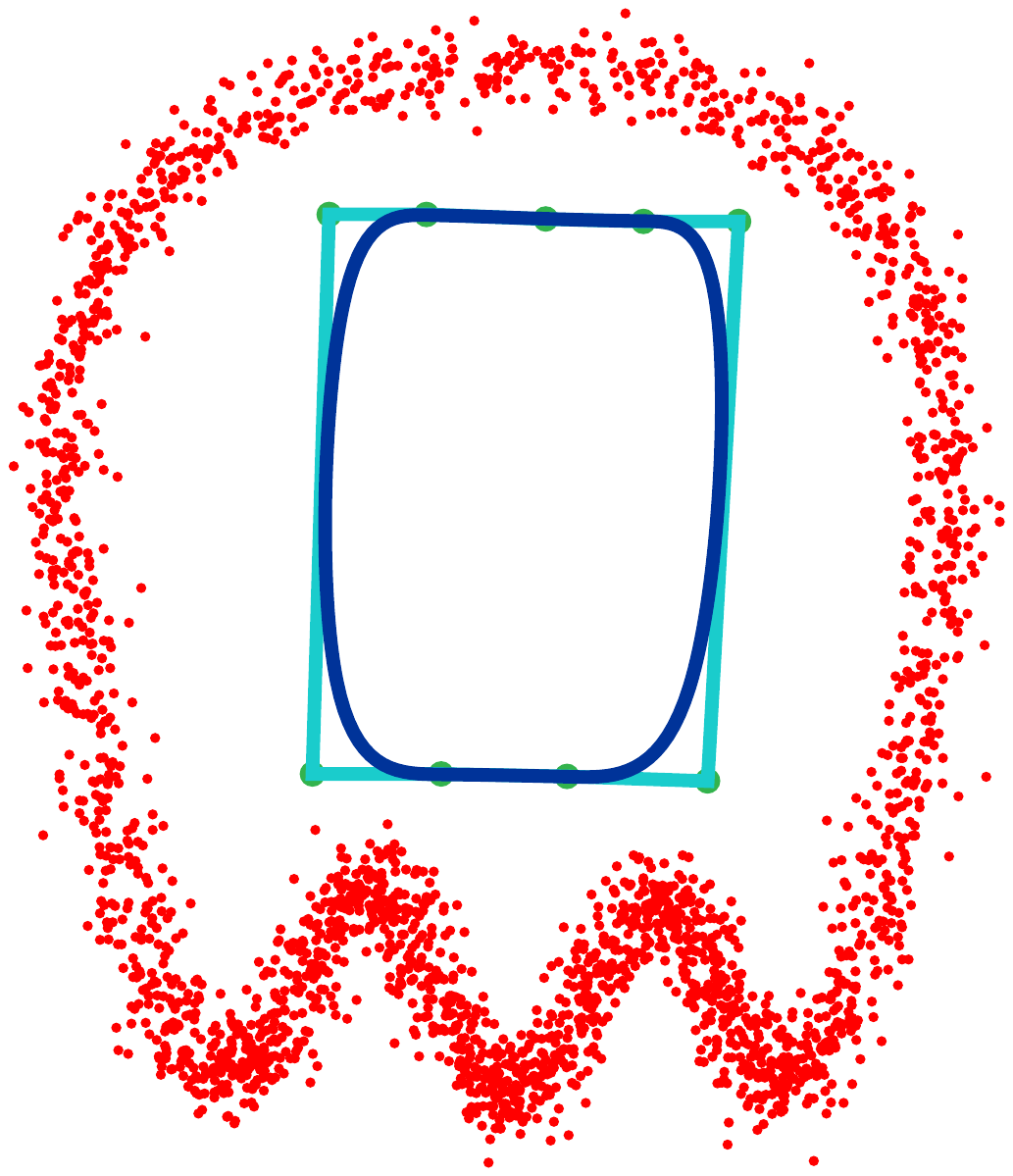}}
\subfigure[Result of L-BFGS.]{\includegraphics[height=2.5cm]{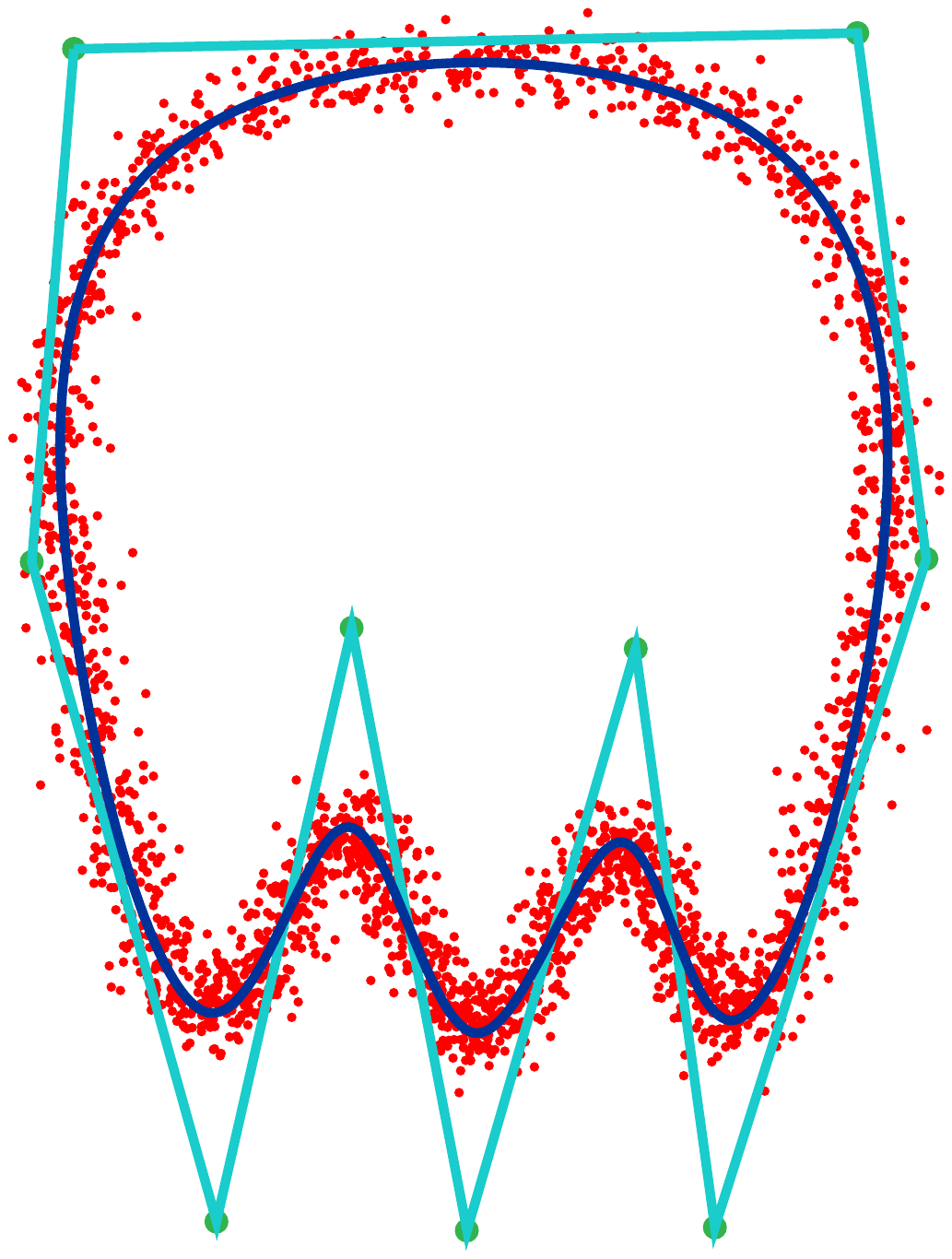}}
\subfigure[Result of SDM.]{\includegraphics[height=2.5cm]{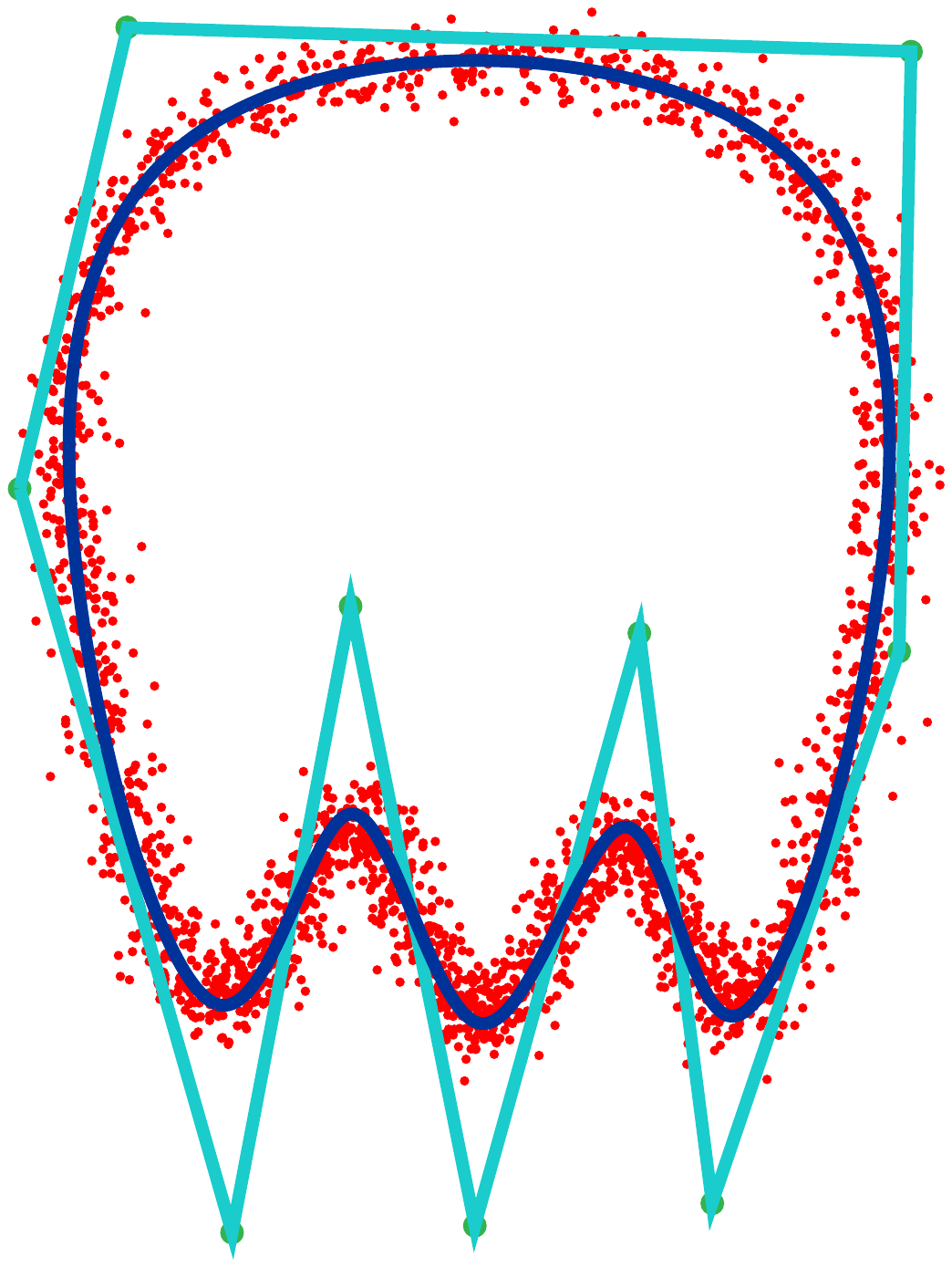}}
\subfigure[Result of Speer's method.]{\includegraphics[height=2.5cm]{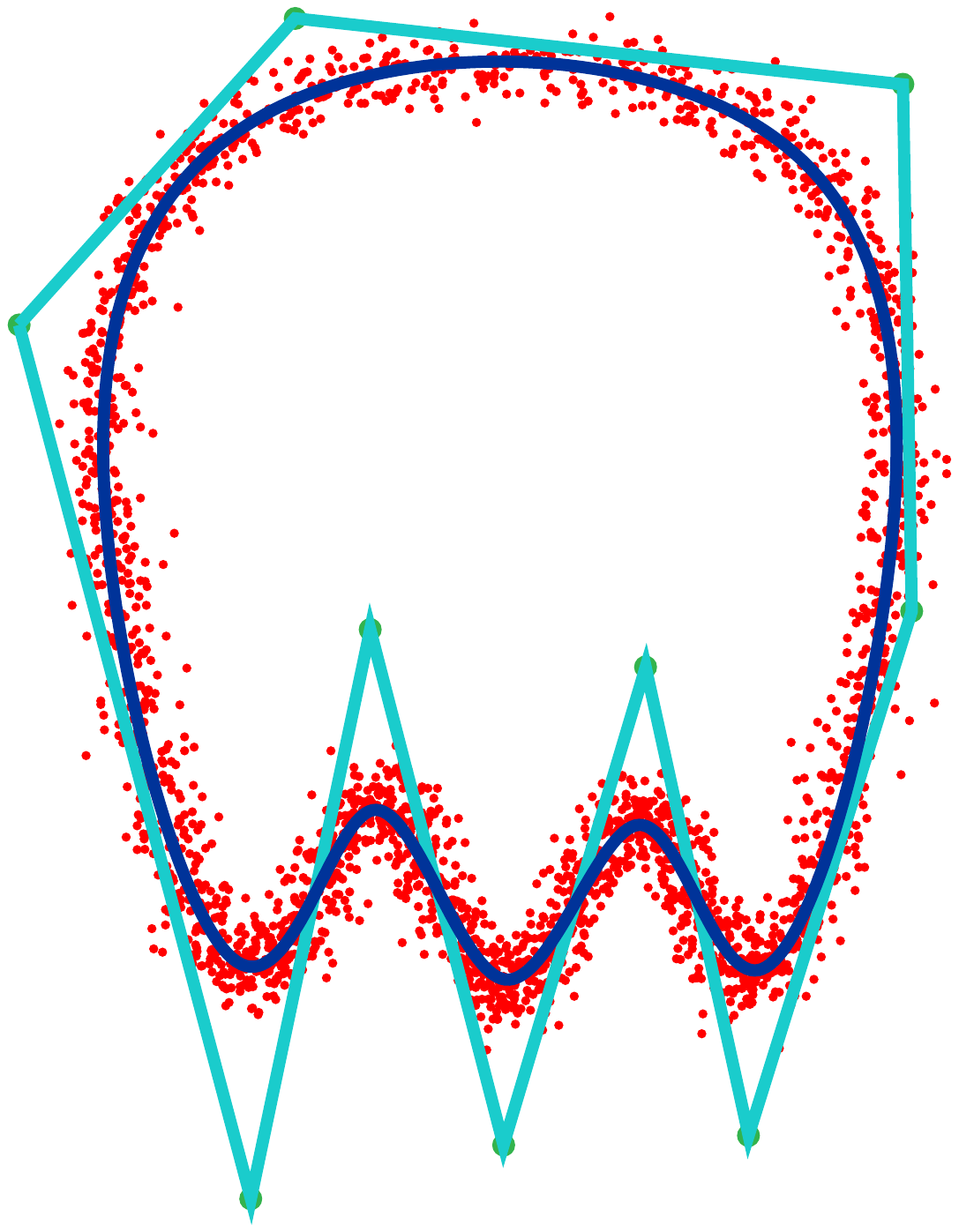}}\\
\vspace{-0.2cm}
\subfigure[Timing and fitting errors.]{
\begin{tabular}{cccc}
\hline
 & L-BFGS & SDM & Speer's \\
 \hline
time(s) & 1.0202 & 9.2754 & 87.5810 \\
error & 0.01796 & 0.01805 & 0.01947 \\
\hline
\end{tabular}
}\\
\vspace{-0.2cm}
\caption{Comparisons with L-BFGS, SDM and Speer's method. The coefficients of fairing term are $\alpha=0$ and $\beta=10^{-3}$. }
\label{fig:speer}
\end{minipage}
\end{figure}

\textbf{Comparison with the method of Speer et al. } In~\cite{Speer1998}, Speer et al proposed to use the Gauss-Newton method to solve the least squares problem~\ref{eq:general}. This method also optimizes $\mathcal{T}$ and $\mathcal{P}$ simultaneously. However, in every iteration it still needs to formulate and solve linear system which includes both location parameters and control points as variables. Therefore, this method is inefficient for large data sets.  Figure~\ref{fig:speer} shows an example with noise containing 2500 data points. We use this example to compare the L-BFGS fitting method with SDM and Speer's method. From Table~\ref{fig:speer} we can see that the L-BFGS fitting method is capable of producing a satisfactory curve about 9 times faster than SDM; SDM in turn is about 8 times faster than Speer's method.

\begin{figure}[t!]
\begin{minipage}{0.45\textwidth}
\centering
\subfigure[Initialization.]{\includegraphics[width=0.45\textwidth]{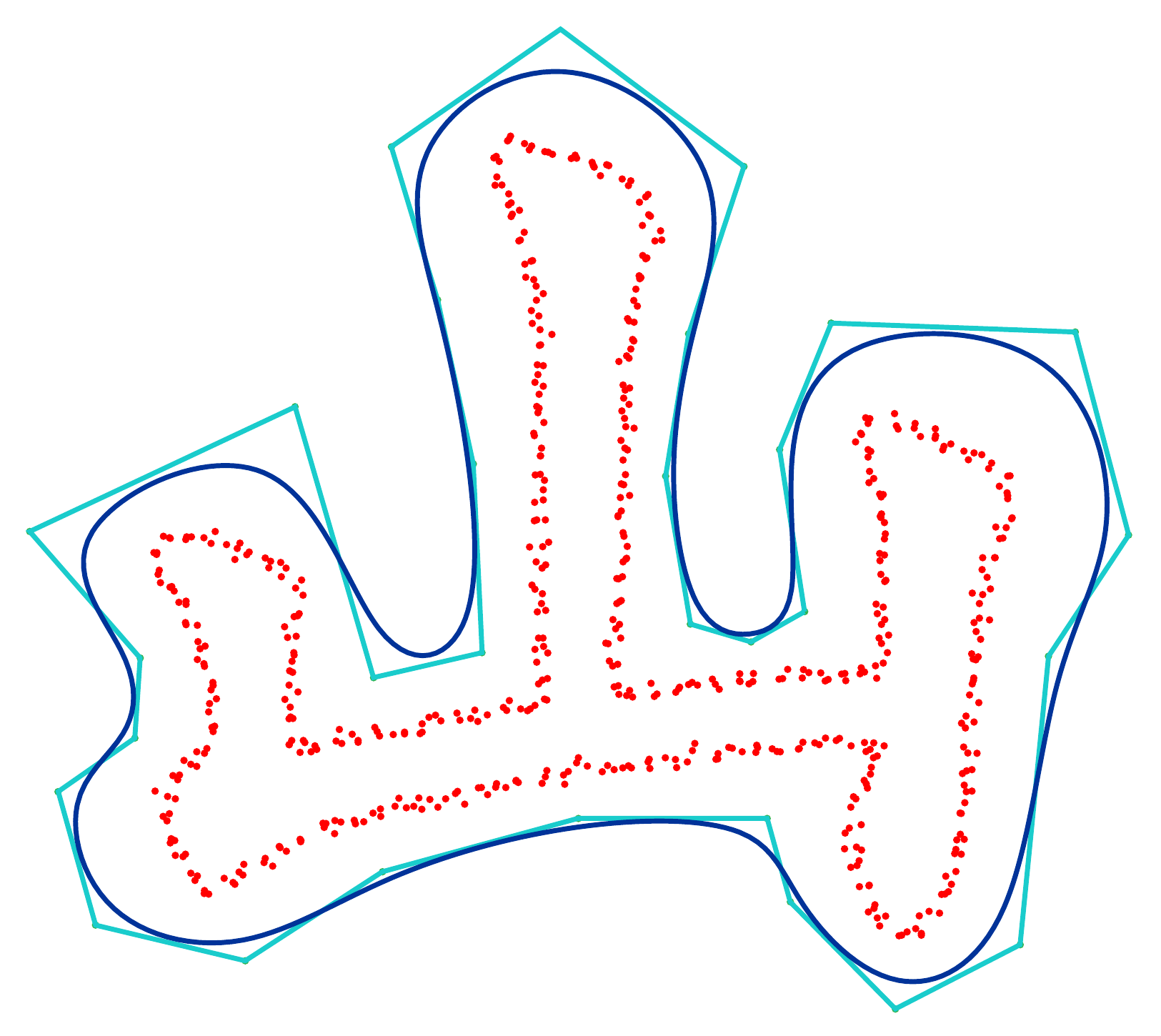}}
\subfigure[Fitting curve.]{\includegraphics[width=0.45\textwidth]{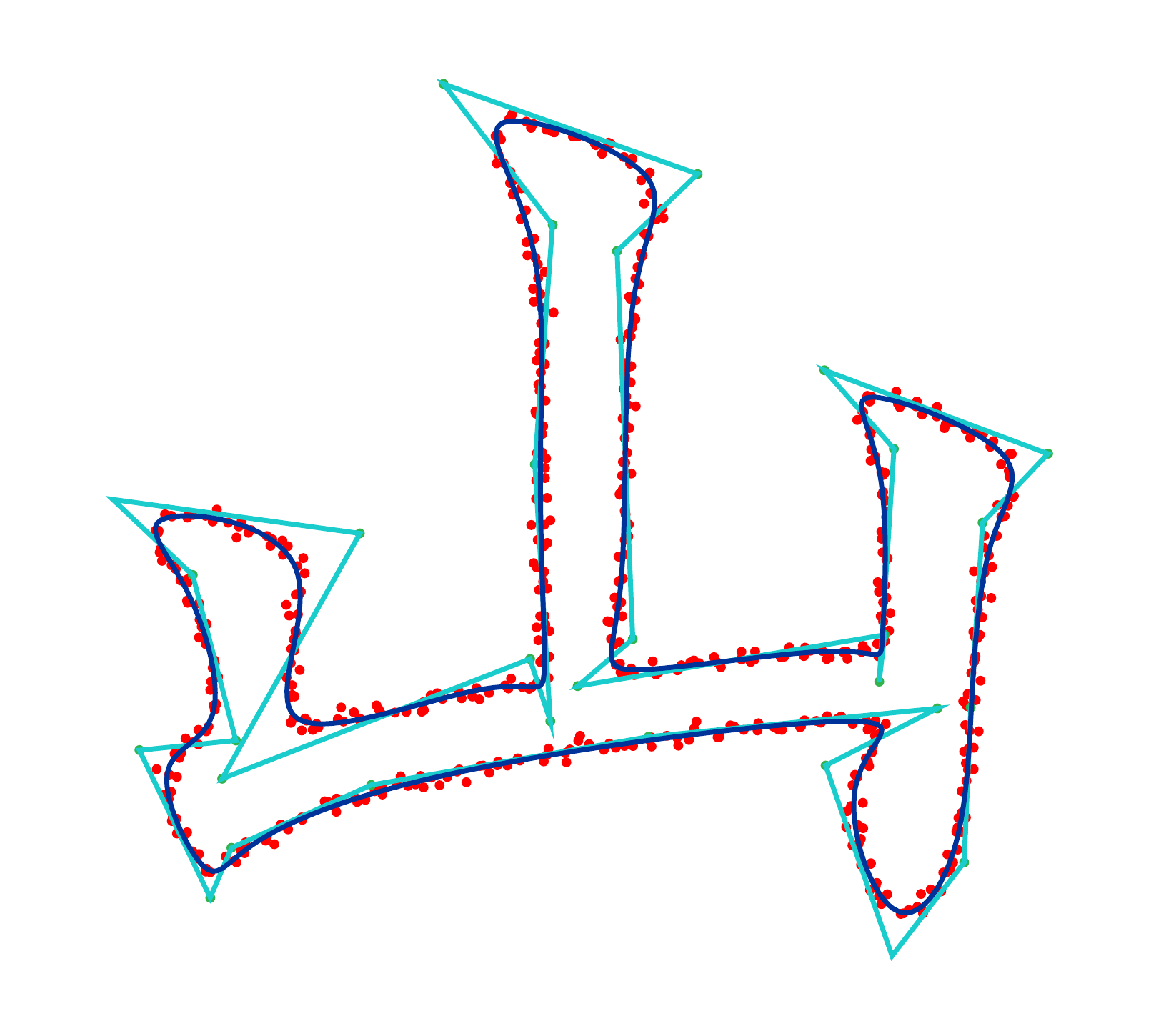}}\\
\vspace{-0.4cm}
\subfigure[Fitting error vs time.]{\includegraphics[width=0.9\textwidth]{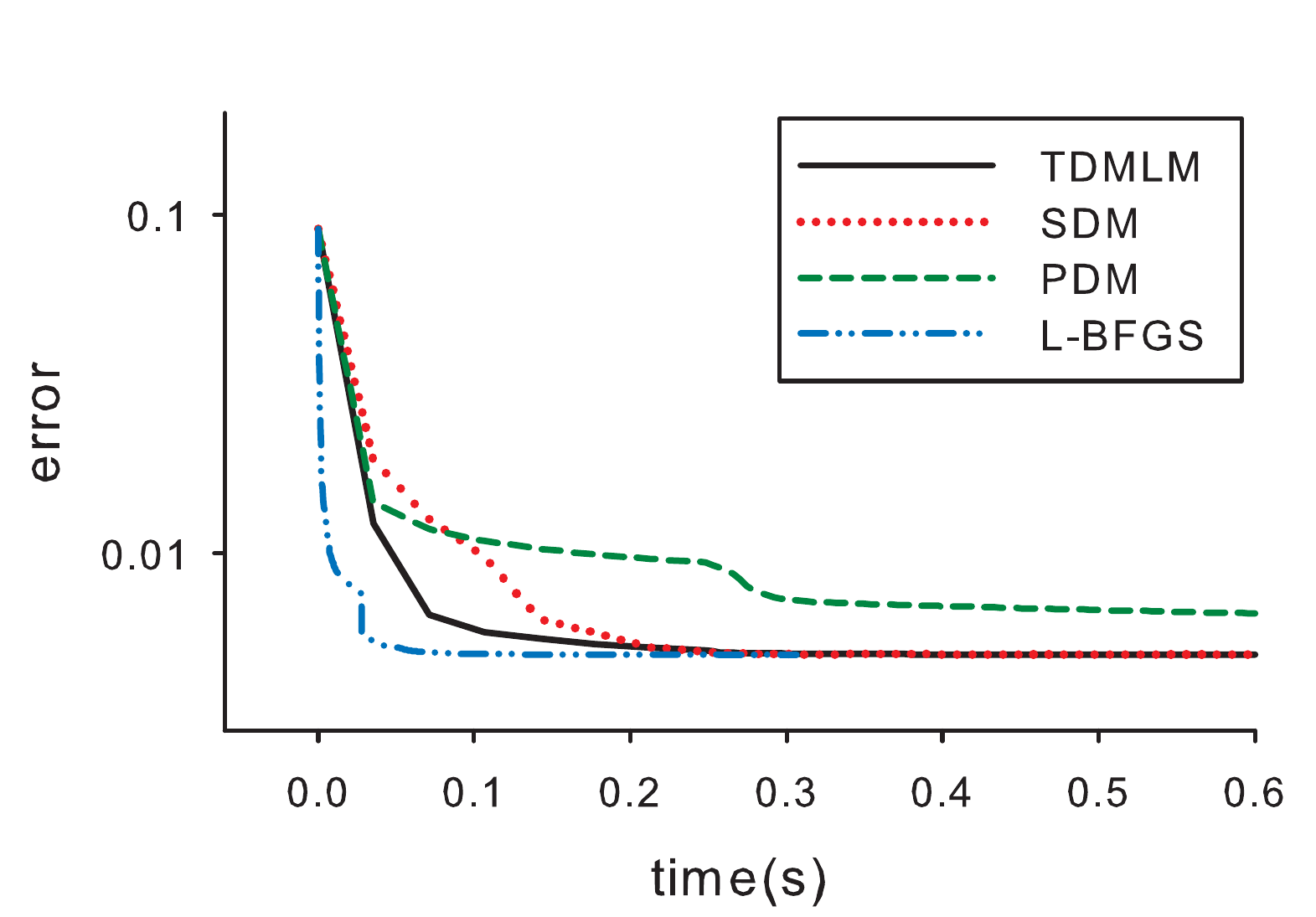}}
\vspace{-0.3cm}
\subfigure[Time to attain minimal error (in seconds).]
{
\begin{tabular}{cccc}
\hline
L-BFGS & PDM & TDMLM & SDM \\
$8.2\cdot 10^{-2}$ & $5.2$ & $0.21$ & $0.23$\\
\hline
\end{tabular}
}\\
\caption{A Chinese character with 30 control points and 600 data points which means "mountain". The coefficients of fairing term are $\alpha=5\cdot 10^{-4}$ and $\beta=0$.}
\label{fig:shan}
\centering
\subfigure[Initialization.]{\includegraphics[width=0.45\textwidth]{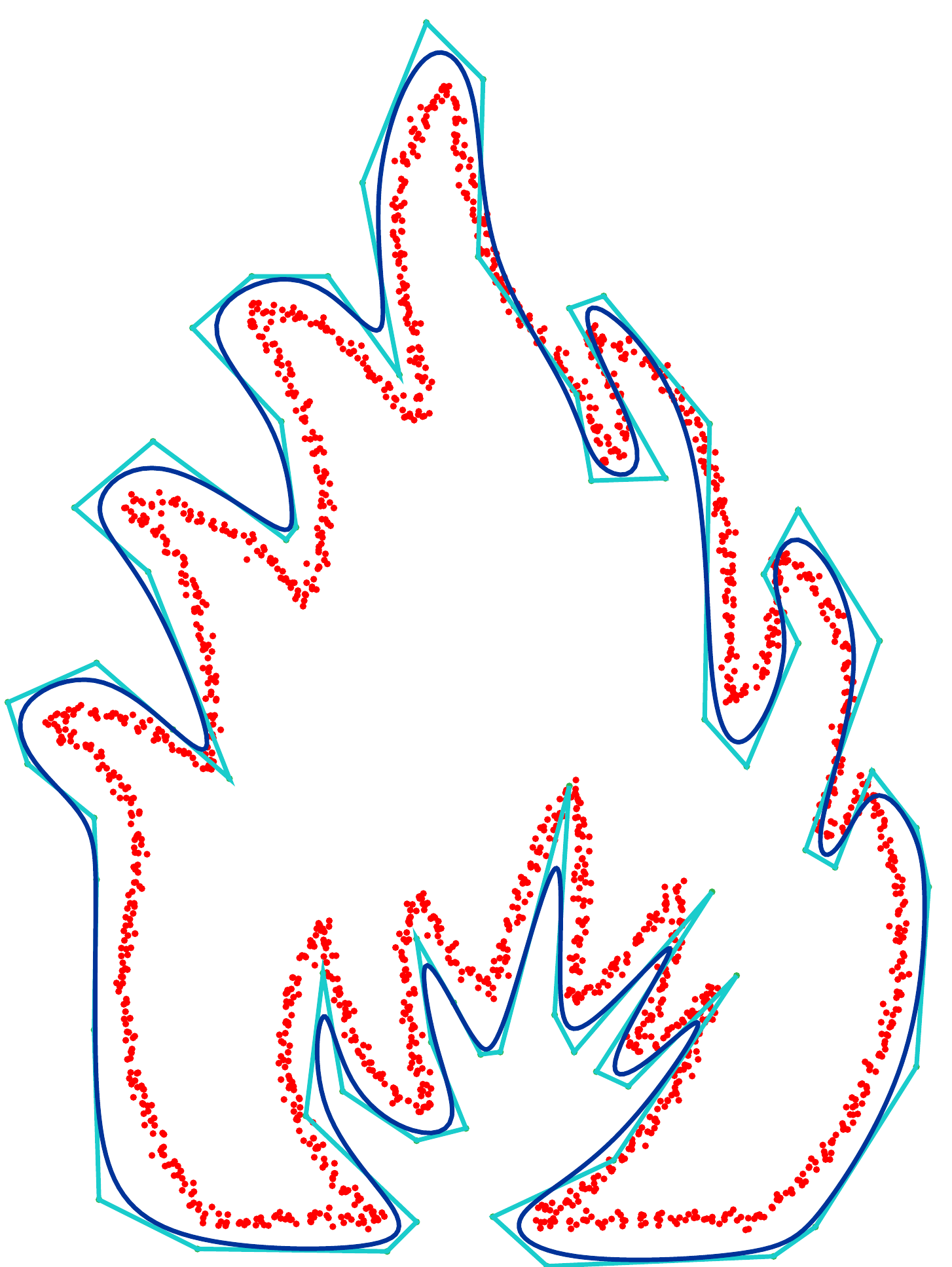}}
\subfigure[Fitting curve.]{\includegraphics[width=0.45\textwidth]{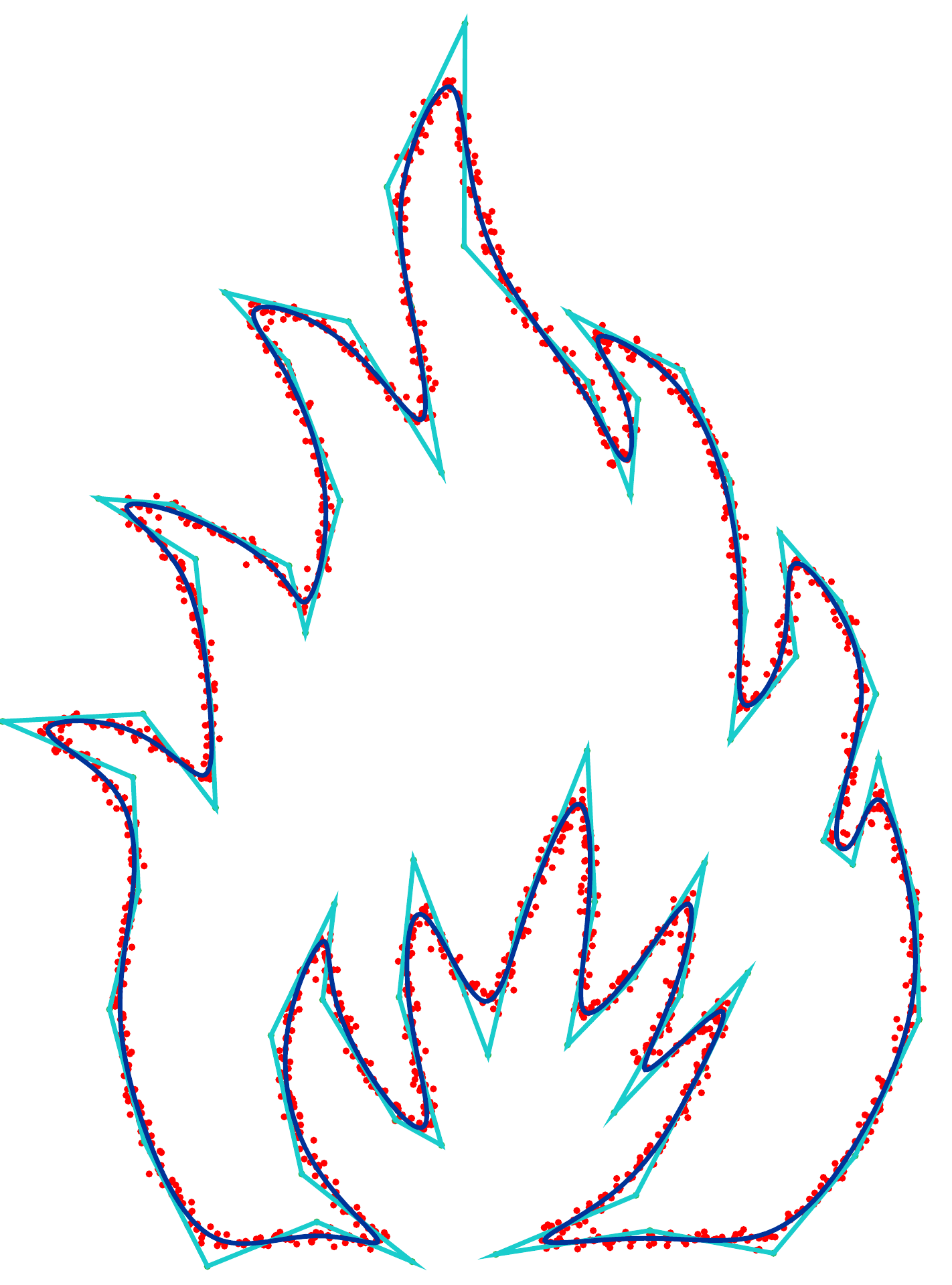}}\\
\vspace{-0.4cm}
\subfigure[Fitting error vs time.]{\includegraphics[width=0.9\textwidth]{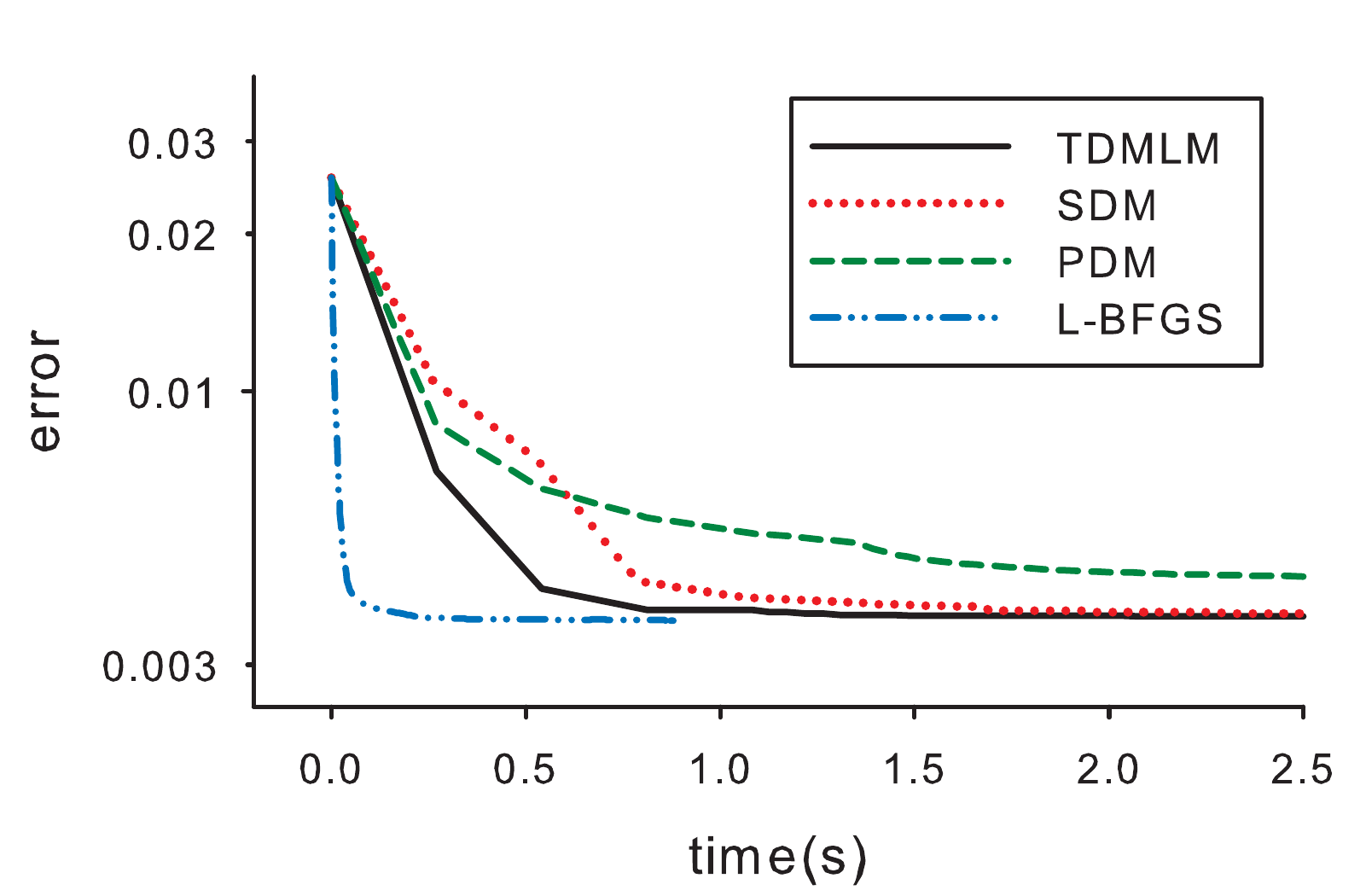}}\\
\vspace{-0.2cm}
\subfigure[Time to attain minimal error (in seconds).]
{
\begin{tabular}{cccc}
\hline
L-BFGS & PDM & TDMLM & SDM \\
$0.28 $ & $41$ & $1.4$ & $1.8$\\
\hline
\end{tabular}
}\\
\caption{Flame with 66 control points and 2000 data points. The coefficients of fairing term are $\alpha=10^{-3}$ and $\beta=10^{-2}$.}
\label{fig:flame}
\end{minipage}
\end{figure}

\textbf{More examples.} We present more examples in Figure~\ref{fig:shan} and \ref{fig:flame}.

\subsection{Analysis and discussions}

\begin{table}[htb]
\centering
\scalebox{0.9}{
\begin{tabular}{cccccc}
\hline
&  &Matrix & Matrix  & foot point \\
&  &filling &solving & projection \\
\hline
Example~\ref{fig:s9} & PDM & 23.8\%  & 15.5\% & 60.6\% \\
(ctrl pts: 6  & TDMLM & 22.6\% & 35.3\% & 41.9\%  \\
data pts: 100)  & SDM & 31.5\% &41.4\% & 26.8\%\\
\hline
Example~\ref{fig:n4} & PDM & 10.7\% & 55.6\% & 33.8\% \\
(ctrl pts: 24 & TDMLM   &7.1\%& 67.3\% & 26.0\% \\
data pts: 90)  & SDM   &8.8\%& 70.0\% & 21.1\% \\
\hline
Example~\ref{fig:9} & PDM & 25.5\% & 26.6\% & 48.3\%\\
(ctrl pts: 8 & TDMLM   &25.5\% & 29.6\% & 45.2\% \\
data pts: 150)  & SDM   &33.9\%&26.6\% & 39.8\% \\
\hline
Example~\ref{fig:shan} & PDM & 21.2\% & 34.2\% & 45.0\%\\
(ctrl pts: 30 &TDMLM       &8.1\% & 78.6\%& 13.0\%\\
data pts: 600) & SDM        &23.1\% &50.2\% &27.2\%\\
\hline
Example~\ref{fig:flame} & PDM  & 9.9\% & 15.1\% & 73.3\%\\
(ctrl pts: 66 & TDMLM  & 15.6\%& 31.8\%& 50.0\%\\
data pts: 2000)  & SDM  & 29.0\%&26.9\%&41.9\%\\
\hline
\end{tabular}
}
\caption{Computational time for different parts of the PDM, TDMLM and SDM methods.}
\label{table:xdmperiter}
\end{table}
\begin{table}[htb]
\centering
\begin{tabular}{cccc}
\hline
 & Computing des- & \multirow{2}{*}{Linesearch}\\
 &cending direction &    \\
\hline
Example~\ref{fig:s9}&95.8\% &4.2\% \\
Example~\ref{fig:n4}&97.8\% & 2.1\% \\
Example~\ref{fig:9} & 96.8\% & 3.2\% \\
Example~\ref{fig:shan}&91.4\%&8.6\%\\
Example~\ref{fig:flame} &89.7\%&10.3\%\\
\hline
\end{tabular}
\caption{Computational time for different parts of the L-BFGS  algorithm. $m$=20.}
\label{table:lbfgsperiter}
\vspace{-0.5cm}
\end{table}

It is difficult to provide a theoretical proof on the superior efficiency of the L-BFGS fitting method over existing methods. As an alternative, in this section we shall conduct an empirical study on the efficiency of PDM, TDMLM, SDM and the L-BFGS fitting method, in order to gain a better understanding of their relative performances. The traditional methods (PDM, TDMLM, SDM) that update control points and location parameters separately mainly include the following tasks: linear system formulation and solving for control points and foot points computation for location parameters. The timing data for different parts of the methods for the examples in this paper are presented in Table~\ref{table:xdmperiter}. The L-BFGS fitting method consists of two parts: the two-loop algorithm for computing a descending direction and a line-search algorithm for deciding step-size. Timings for these two parts on the same examples as in Table~\ref{table:xdmperiter} are listed in Table~\ref{table:lbfgsperiter}.

We have the following observations on these timing data.

\begin{itemize}
\item Although the number of control points is generally much fewer than the number of data points, traditional methods still consume more than half of the total time on updating control points, because of the need to fill the matrix and solving the linear system in every iteration. The L-BFGS algorithm (Algorithm~\ref{lbfgsalgorithm}) is a Newton-type optimization method that uses an approximated inverse Hessian matrix of the objective function. However, instead of solving a large linear system to compute the descend direction as PDM, TDMLM and SDM, the L-BFGS algorithm uses a two-loop algorithm which uses only vector multiplications and is therefore much faster.

\item Foot point computation is very time consuming.  If we re-compute the initialization of foot points in every iteration, the overall time for foot point computation would be more than 90$\%$ of the total time of the algorithm, as observed in~\cite{Wang2006}. In our implementation of the traditional methods used for comparison in this paper, we use as much as possible the foot points in the previous iteration as initialization for the current iteration, thus having saved a lot of time for traditional methods. Even so, the L-BFGS fitting methods still outperforms these traditional methods, since there is generally no need to perform foot point projection in the L-BFGS fitting method. In rare cases, foot point computation is needed for the L-BFGS method to jump out of a poor local minimum, as we have explained in section 4. This is an issue mostly due to the quality of initialization, rather than the inherent demand of the algorithm.

\item The L-BFGS fitting method  performs optimization in a much higher dimensional space than those of traditional methods since generally the number of data points is much larger than that of the control points. Therefore, the terrain of the functional is supposed to be much more complicated and the optimization is more difficult. The linesearch algorithm is  therefore necessary for stable convergence of the L-BFGS fitting algorithm. Table~\ref{table:lbfgsperiter} shows that the computational time by the linesearch algorithm usually takes a small part of the total time (less than 10$\%$ in most cases).
\end{itemize}

\begin{figure}[t!]
\begin{minipage}{0.45\textwidth}
\centering
\subfigure[100 data points.]{\includegraphics[height=1.8cm]{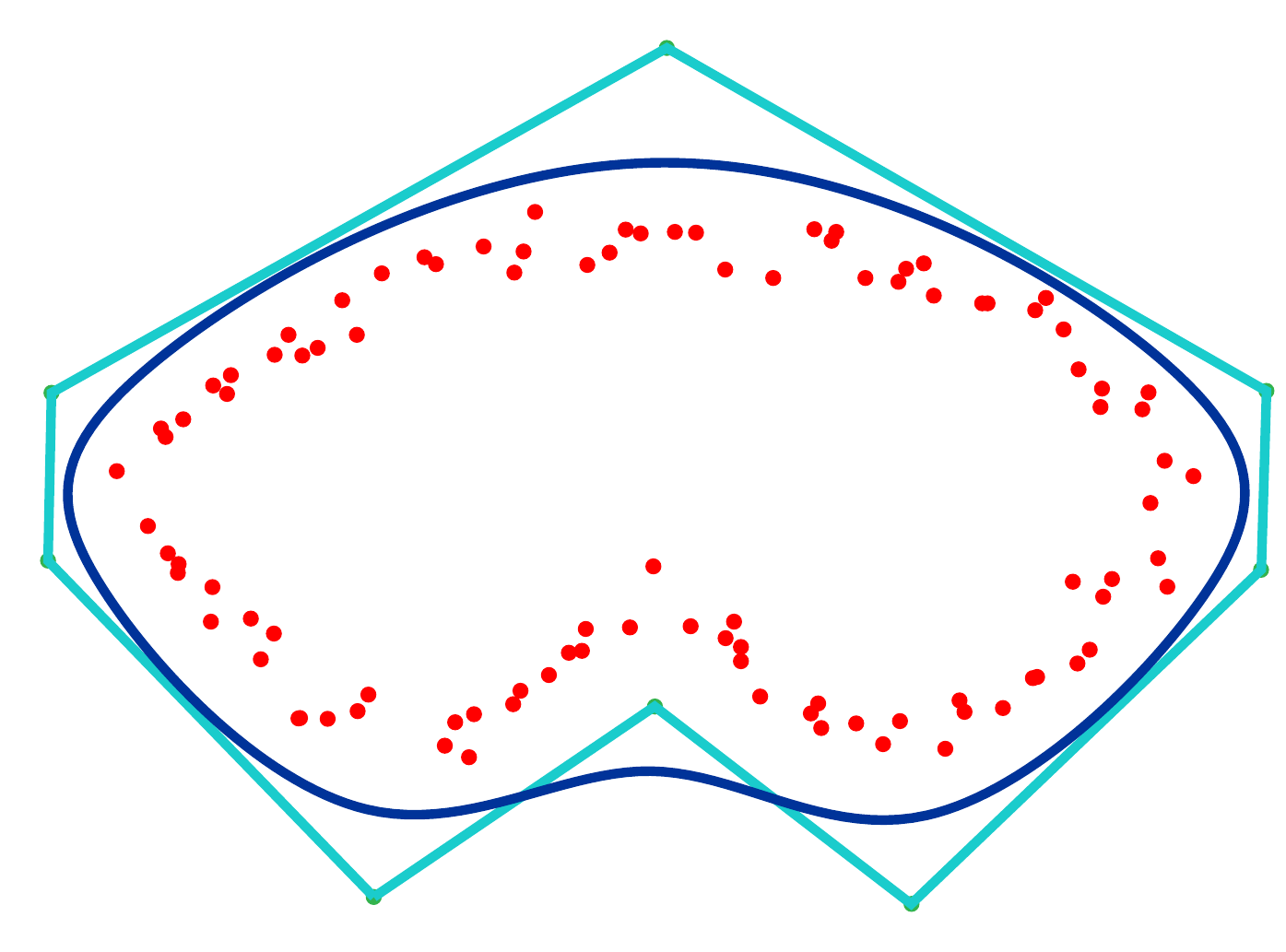}}
\subfigure[200 data points.]{\includegraphics[height=1.8cm]{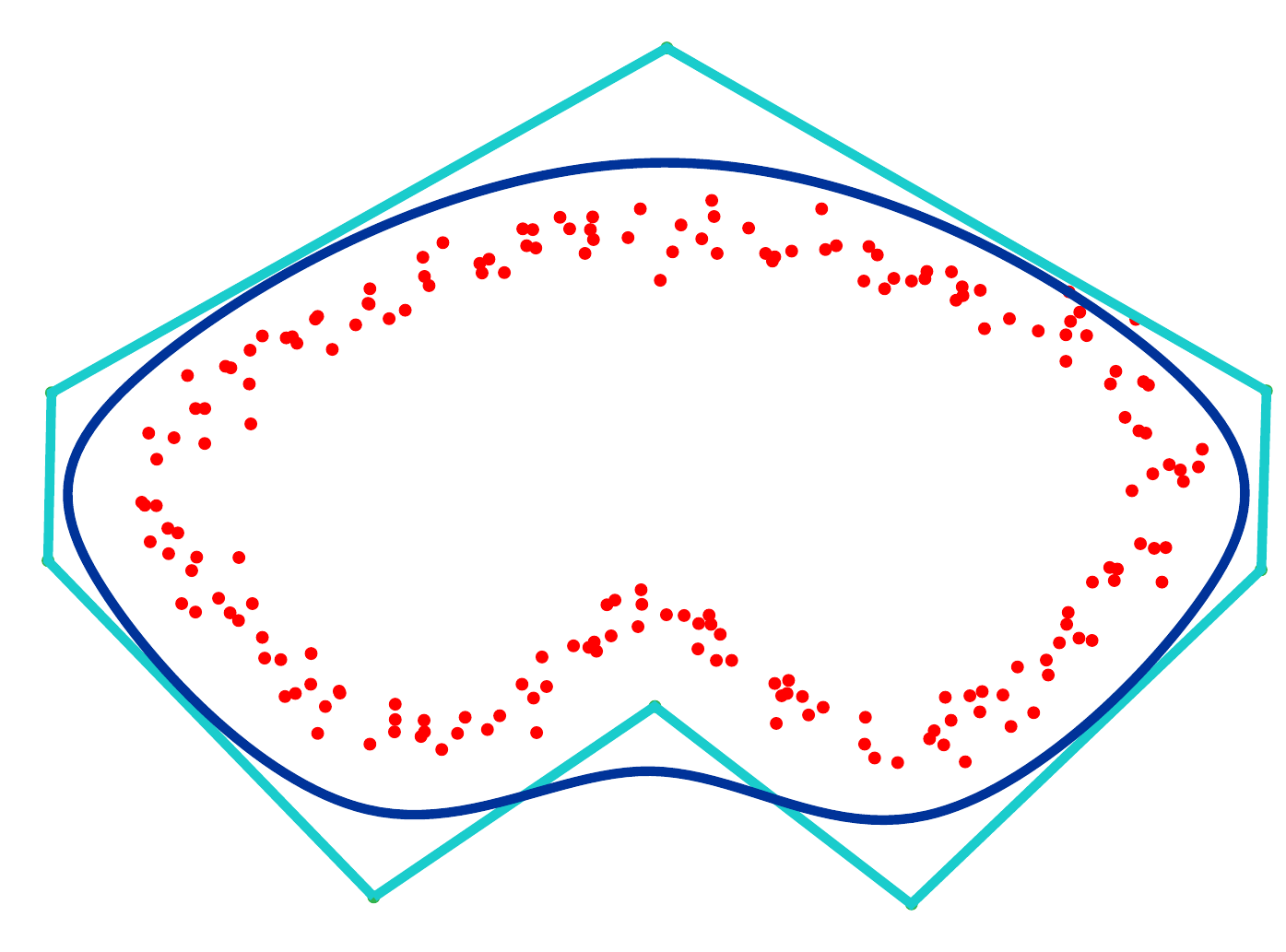}}
\subfigure[500 data points.]{\includegraphics[height=1.8cm]{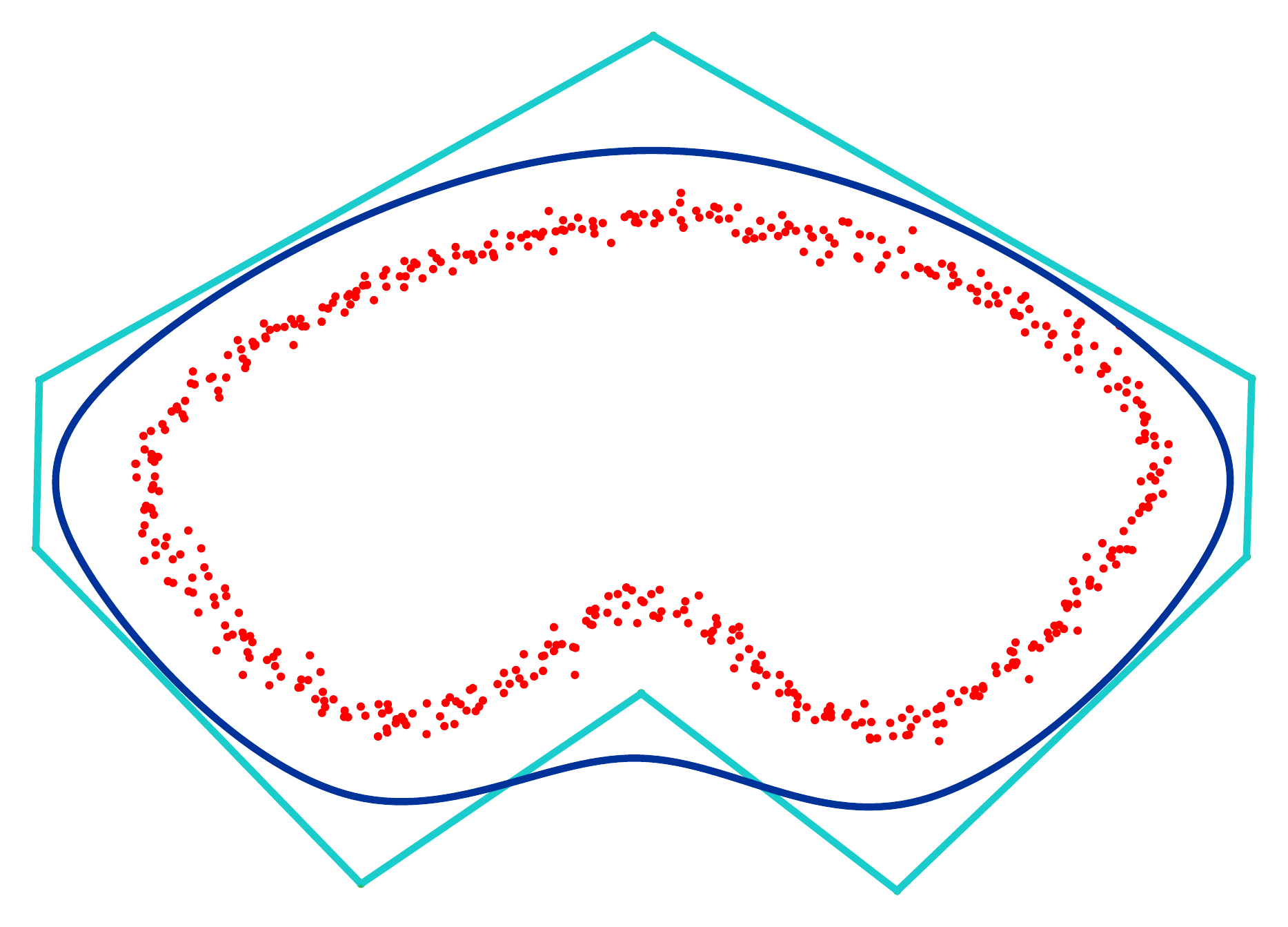}}\\
\subfigure[1000 data points.]{\includegraphics[height=1.8cm]{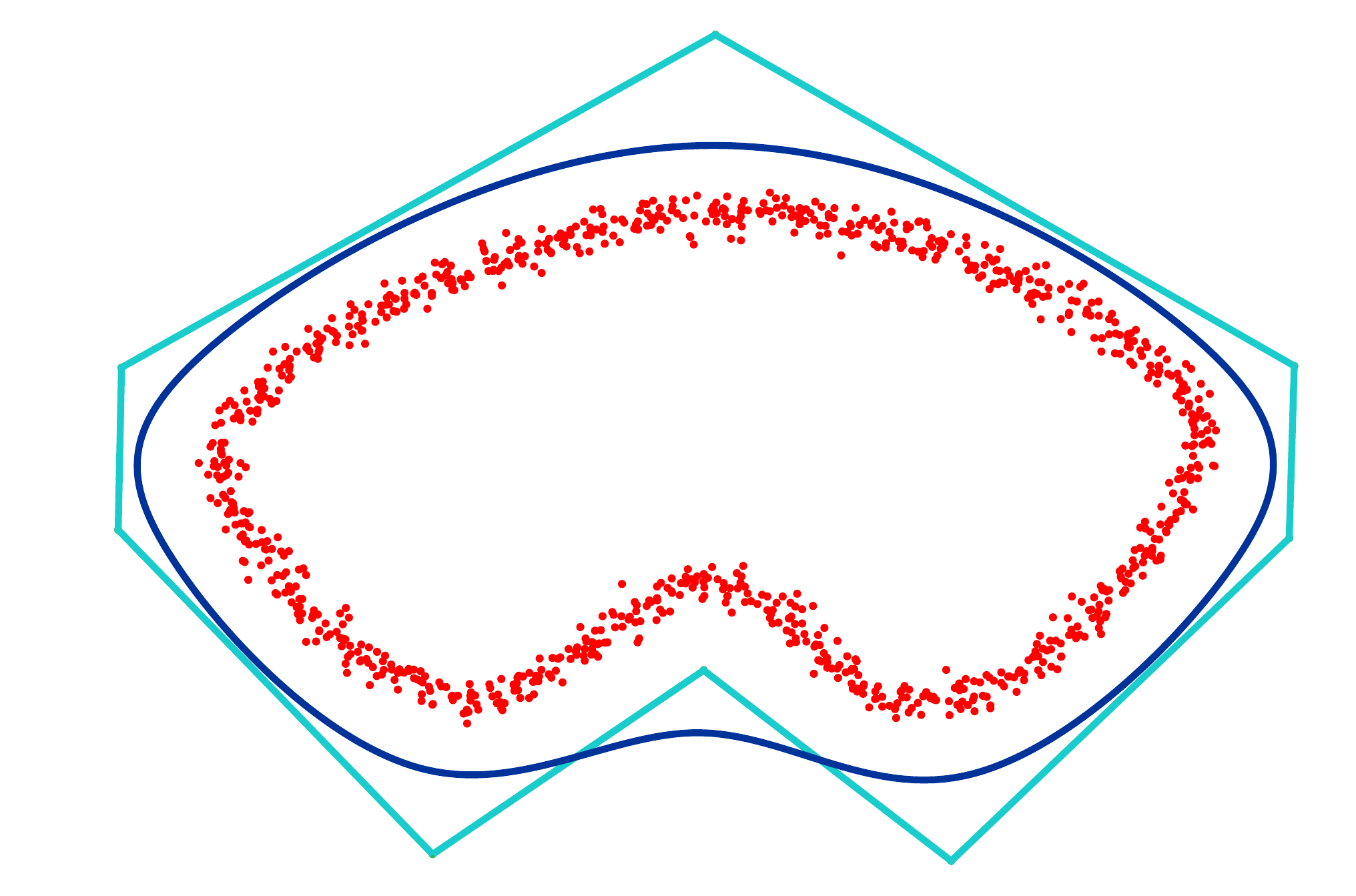}}
\hspace{0.3cm}
\subfigure[3000 data points.]{\includegraphics[height=1.8cm]{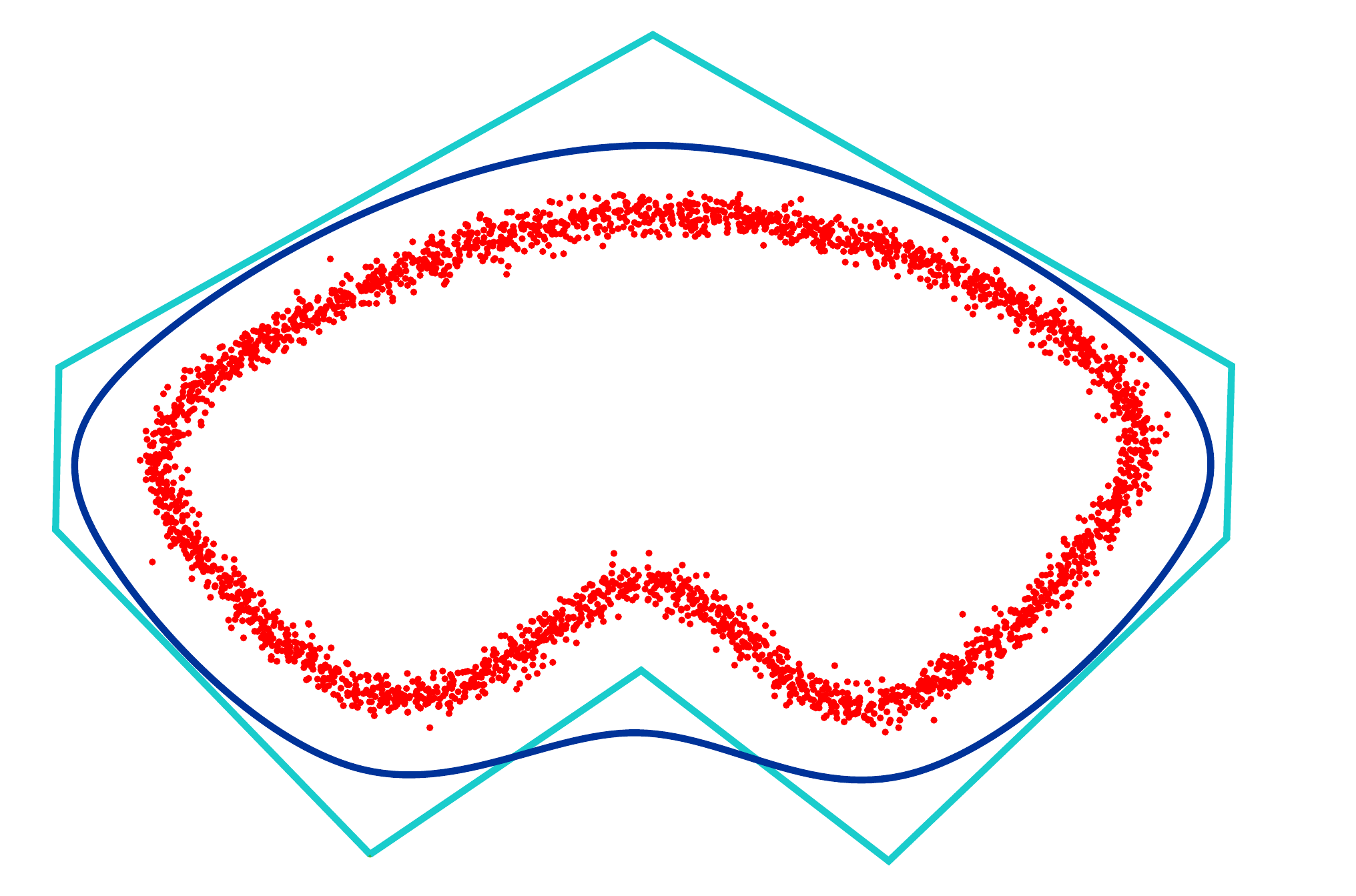}}\\
\vspace{-0.3cm}
\subfigure[Per-iteration time as the number of data points increases.]{\includegraphics[width=0.99\textwidth]{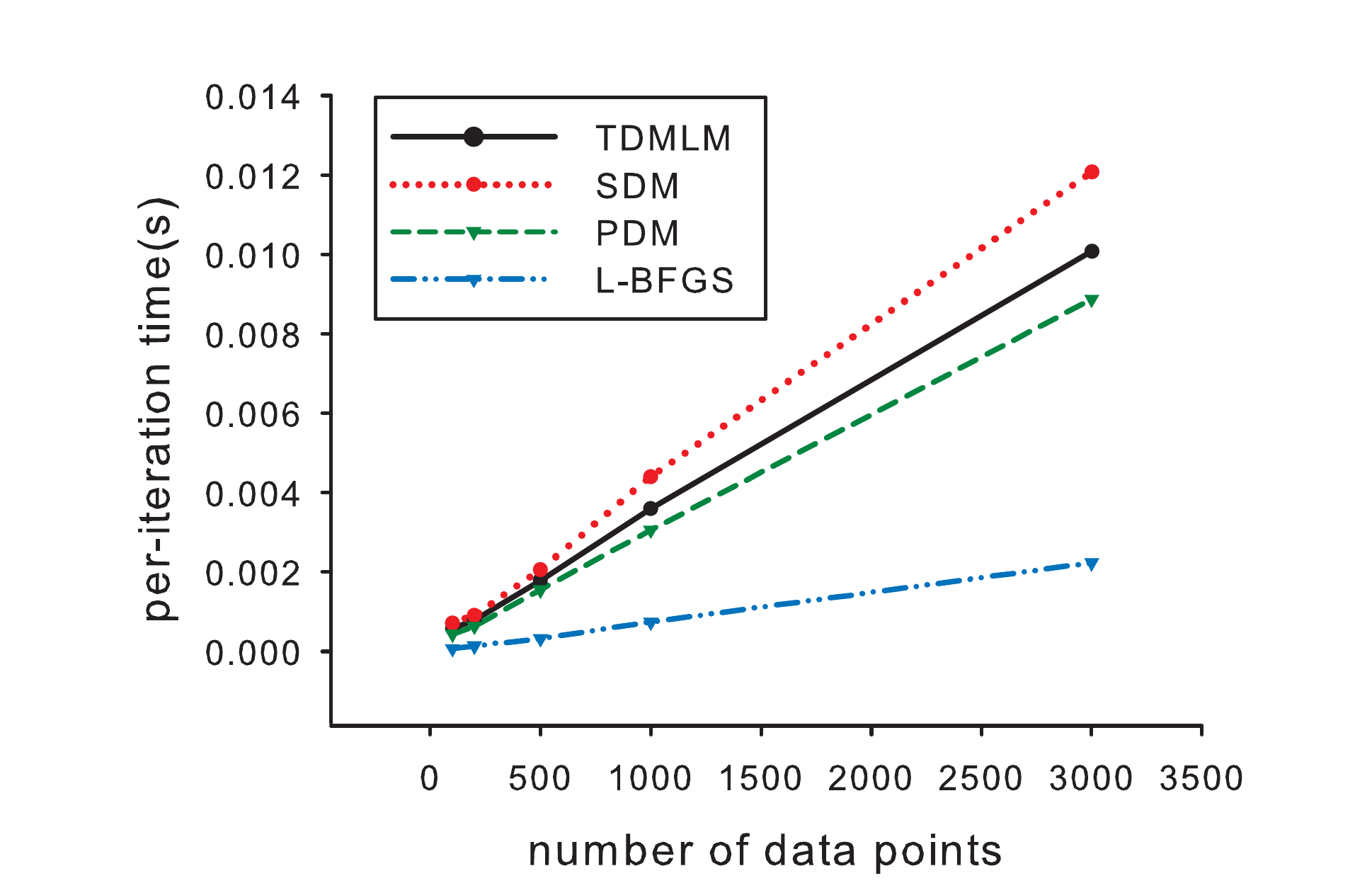}}
\vspace{-0.2cm}
\caption{Increasing data points: An example with 8 control points. The coefficients of fairing term are $\alpha=5\cdot 10^{-4}$ and $\beta=0$}
\label{fig:noisydataincrease}
\vspace{0.4cm}
\centering
\subfigure[10 Control points]{\includegraphics[height=1.6cm]{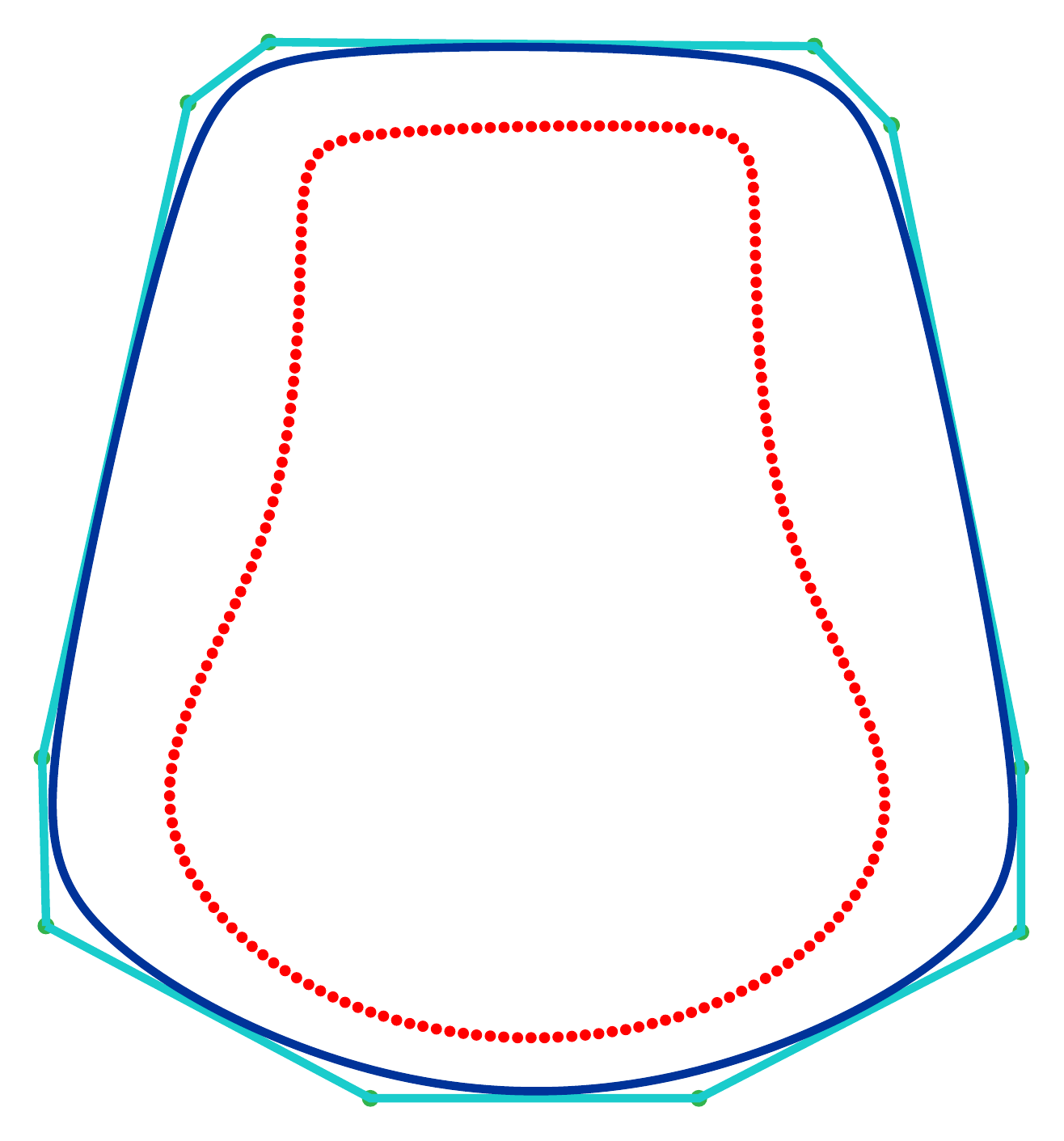}}
\hspace{0.3cm}
\subfigure[20 Control points]{\includegraphics[height=1.6cm]{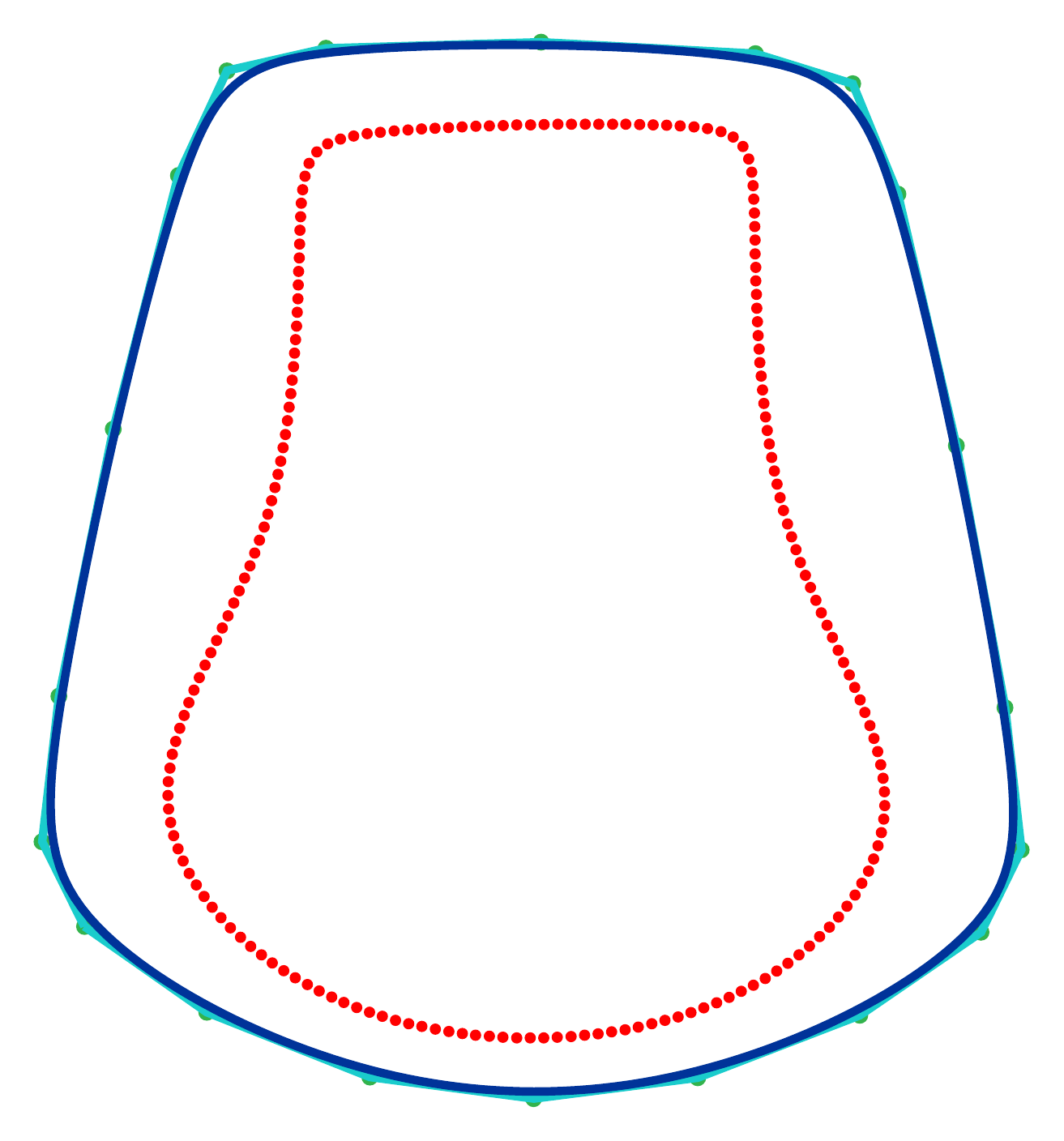}}
\hspace{0.3cm}
\subfigure[40 Control points]{\includegraphics[height=1.6cm]{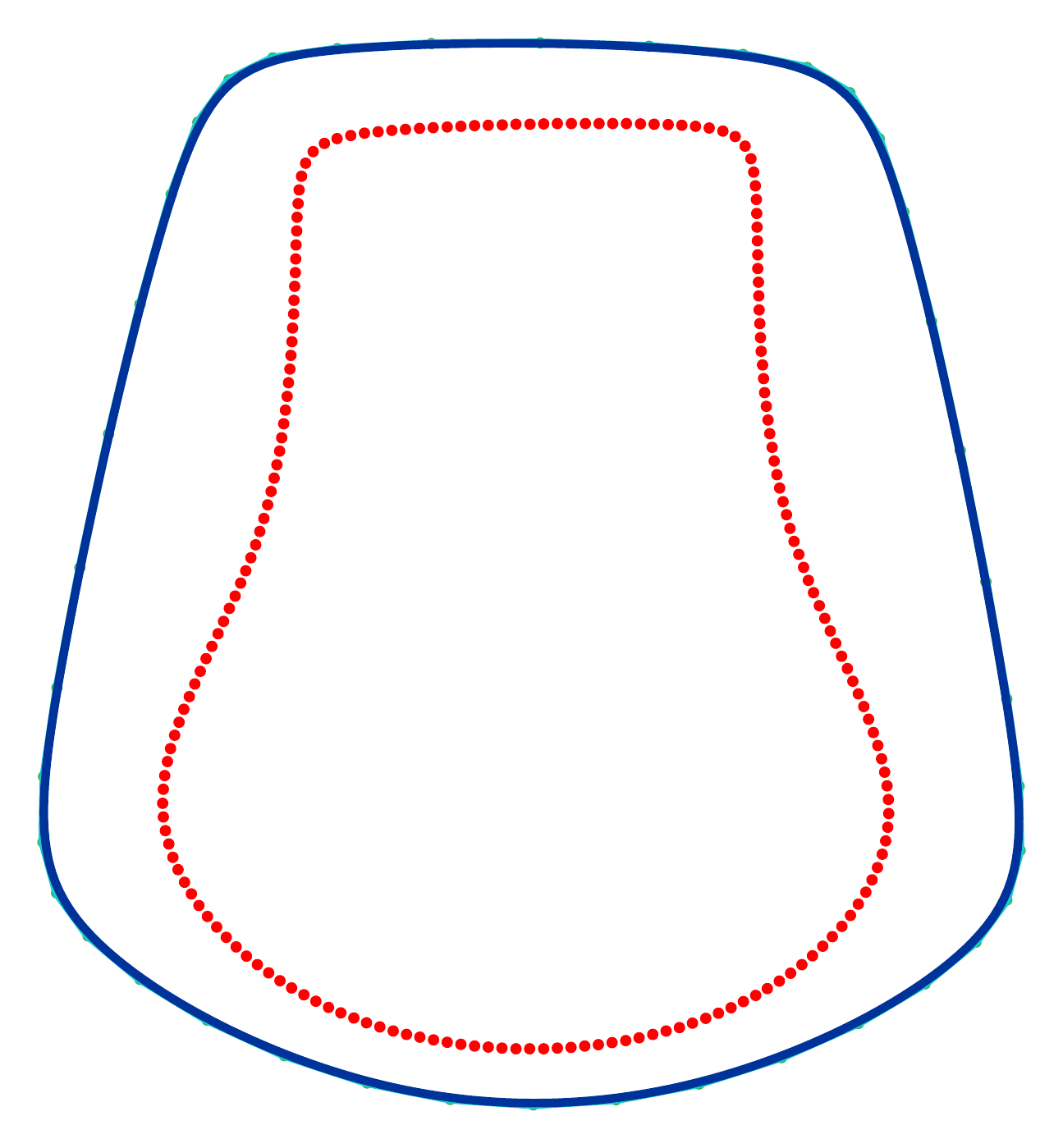}}
\hspace{0.3cm}
\subfigure[80 Control points]{\includegraphics[height=1.6cm]{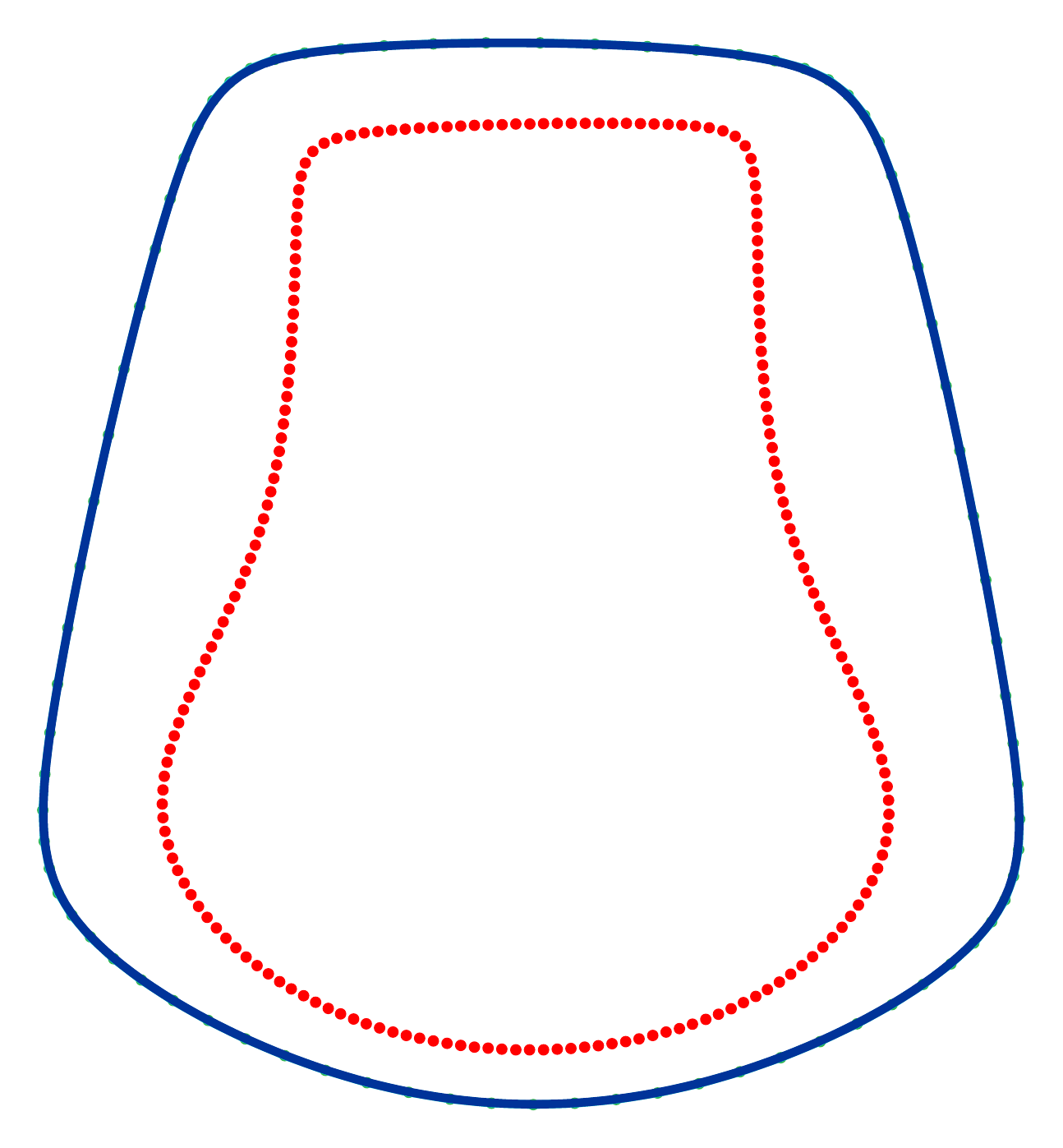}}\\
\vspace{-0.3cm}
\subfigure[Per-iteration time as the number of control points increases.]{\includegraphics[width=0.99\textwidth]{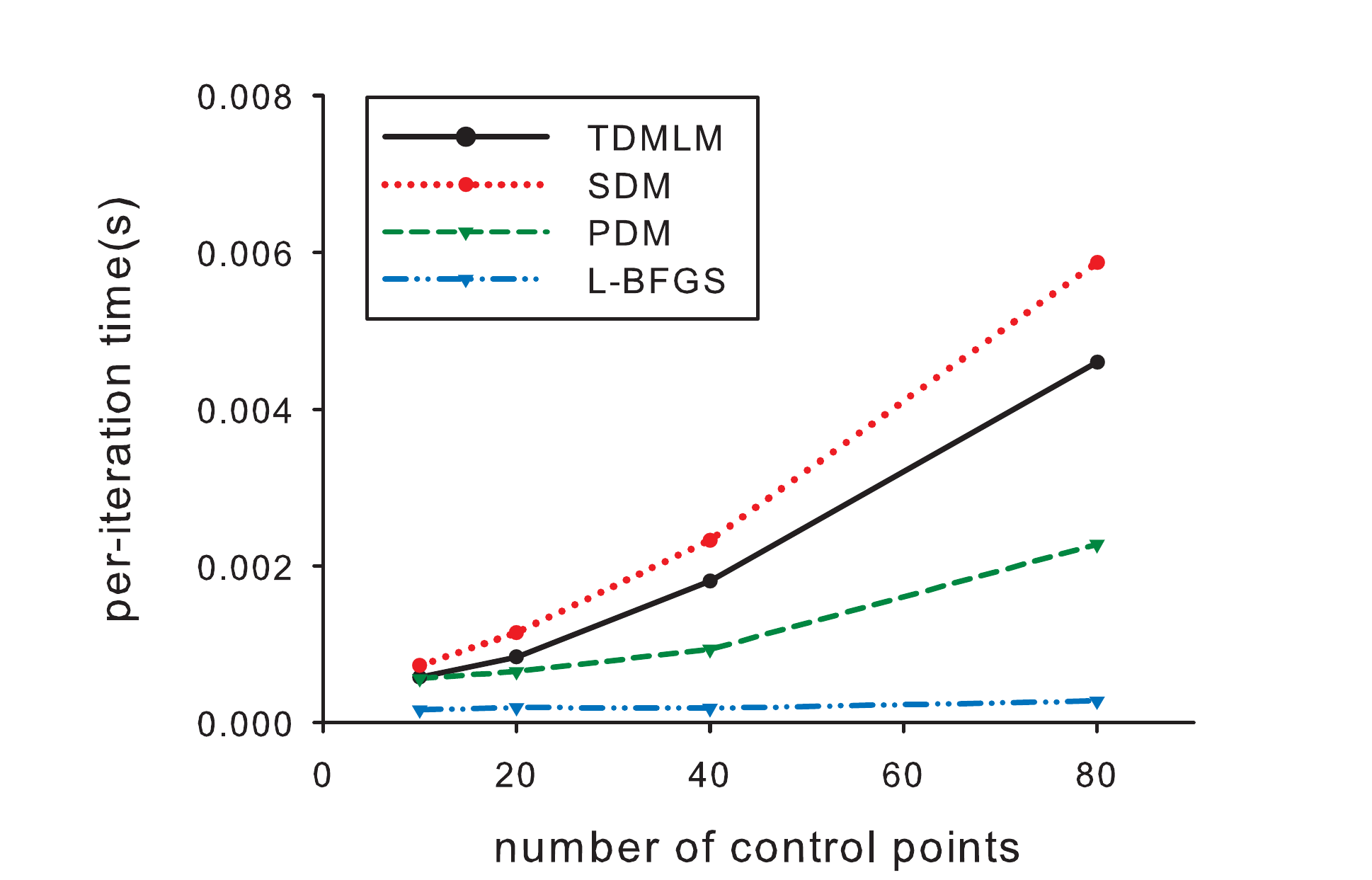}}
\vspace{-0.2cm}
\caption{With increased number of control points. There are 200 data points. No fairing term is used.}
\label{fig:smoothctrlptsincrease}
\vspace{-0.3cm}
\end{minipage}

\end{figure}

We now study how the computational time depends on the number of control points and and the number of data points.

{\bf Timing vs \# of data points}.
In Figure~\ref{fig:noisydataincrease}, we show computational time with increased number of data points for various methods.  The  number of data points in these point sets is 100, 200, 500, 1000 and 3000 respectively. The fitting curve has 8 control points. Figure~\ref{fig:noisydataincrease} shows that the computational time for each iteration of all 4 methods depends almost linearly on  the number of data points.
This can be explained as follows. It is not difficult to see that in the PDM, TDMLM and SDM, the time for  matrix building and foot point projection is linear in the number of data points. The time for solving linear system is constant since the number of control points is fixed. Consequently, the total time for these three methods increase linearly as the number of data points increases. For the L-BFGS fitting method, computational time is linear in the number of variables (2 $\times$ the number of control points $+$ the number of data points), therefore the computational time of the L-BFGS fitting method also increases  linearly as the number of data points increases.

{\bf Timing vs \# of control points}. The  relationship of computational time and the number of control points of the fitting B-spline curve can be observed in Figure~\ref{fig:smoothctrlptsincrease}. We insert new control points by knot insertion in each knot interval and get 4 B-spline curves with the number of control points: 10, 20, 40 and 80 respectively. The number of target data points is 200. We see that the per-iteration time for the tested traditional methods increases faster than the L-BFGS fitting method when the number B-spline control points increases.  That is because in the PDM, TDMLM and SDM, the size of linear system is quadratic to the number of control points, but the computational time of the L-BFGS algorithm (i.e. the two-loop algorithm and the linesearch) depends on the number of control points linearly.

These experiments show that the L-BFGS fitting method is more suitable for large scale curve fitting problems, especially when the target shape is complicated and a large number of control points are involved.

\section{Conclusion and Future Work}

In this paper, we propose a new curve fitting method based on the L-BFGS optimization technique. The unique features of this algorithm are that it does not need to perform the time-consuming foot point projection in every iteration as in traditional approaches and that it does not need to formulate and solve a linear system of equations in every iteration; instead, it uses only efficient vector multiplications. As a result, this new method is much faster than other traditional methods, as demonstrated by a number of experimental results presented. In the future we will extend this method to solving the B-spline surface fitting problem, for which we expect even more significant improvements over the existing methods because of the large number of data points as well as the large number of control points involved in surface fitting.


\bibliographystyle{abbrv}
\bibliography{draft}

\begin{thebibliography}{10}

\bibitem{jue:05e}
M.~Aigner and B.~J\"uttler.
\newblock Robust computation of foot points on implicitly defined curves.
\newblock In M.~D\ae{}hlen, K.~M\o{}rken, and L.~Schumaker, editors, {\em
  Mathematical Methods for Curves and Surfaces: Troms\o{} 2004}, pages 1--10.
  Nashboro Press, Brentwood, 2005.

\bibitem{Alhanaty2001a}
M.~Alhanaty and M.~Bercovier.
\newblock {Curve and surface fitting and design by optimal control methods}.
\newblock {\em Computer-Aided Design}, 33(2):167--182, 2001.

\bibitem{Bjorck1996}
{\AA}.~Bj{\"o}rck.
\newblock {\em {Numerical methods for least squares problems}}.
\newblock Society for Industrial Mathematics, 1996.

\bibitem{Blake1998}
A.~Blake and M.~Isard.
\newblock {\em {Active contours}}, volume~2.
\newblock Springer London, 1998.

\bibitem{Forsey:1995:SFH:221659.221665}
D.~R. Forsey and R.~H. Bartels.
\newblock Surface fitting with hierarchical splines.
\newblock {\em ACM Trans. Graph.}, 14:134--161, April 1995.

\bibitem{haber2001smooth}
J.~Haber, F.~Zeilfelder, O.~Davydov, and H.~Seidel.
\newblock Smooth approximation and rendering of large scattered data sets.
\newblock In {\em Visualization, 2001. VIS'01. Proceedings}, pages 341--571.
  IEEE.

\bibitem{Hoppe:1994:PSS:192161.192233}
H.~Hoppe, T.~DeRose, T.~Duchamp, M.~Halstead, H.~Jin, J.~McDonald,
  J.~Schweitzer, and W.~Stuetzle.
\newblock Piecewise smooth surface reconstruction.
\newblock In {\em Proceedings of the 21st annual conference on Computer
  graphics and interactive techniques}, SIGGRAPH '94, pages 295--302, New York,
  NY, USA, 1994. ACM.

\bibitem{Hoschek1988}
J.~Hoschek.
\newblock {Intrinsic parametrization for approximation}.
\newblock {\em Computer Aided Geometric Design}, 5(1):27--31, 1988.

\bibitem{Hoschek:1989:OAC:80732.80734}
J.~Hoschek, F.-J. Schneider, and P.~Wassum.
\newblock Optimal approximate conversion of spline surfaces.
\newblock {\em Comput. Aided Geom. Des.}, 6:293--306, October 1989.

\bibitem{Hu2005}
S.-M. Hu and J.~Wallner.
\newblock A second order algorithm for orthogonal projection onto curves and
  surfaces.
\newblock {\em Comput. Aided Geom. Des.}, 22:251--260, March 2005.

\bibitem{jiang2004preconditioned}
L.~Jiang, R.~Byrd, E.~Eskow, and R.~Schnabel.
\newblock {A preconditioned L-BFGS algorithm with application to molecular
  energy minimization}.
\newblock Technical report, Colorado Univ at Boulder Dept of Computer Science,
  2004.

\bibitem{Liu2008}
Y.~Liu and W.~Wang.
\newblock {A revisit to least squares orthogonal distance fitting of parametric
  curves and surfaces}.
\newblock In {\em Proceedings of the 5th international conference on Advances
  in geometric modeling and processing}, pages 384--397, Hangzhou, China, 2008.
  Springer-Verlag.

\bibitem{Nocedal1999}
J.~Nocedal and S.~J. Wright.
\newblock {\em {Numerical optimization}}.
\newblock Springer verlag, 1999.

\bibitem{Pavlidis1983}
T.~Pavlidis.
\newblock {Curve fitting with conic splines}.
\newblock {\em ACM Transactions on Graphics (TOG)}, 2(1):1--31, 1983.

\bibitem{Plass:1983:CPP:800059.801153}
M.~Plass and M.~Stone.
\newblock Curve-fitting with piecewise parametric cubics.
\newblock In {\em Proceedings of the 10th annual conference on Computer
  graphics and interactive techniques}, SIGGRAPH '83, pages 229--239, New York,
  NY, USA, 1983. ACM.

\bibitem{Pottmann2003}
H.~Pottmann and M.~Hofer.
\newblock {Geometry of the squared distance function to curves and surfaces}.
\newblock {\em Visualization and mathematics III}, pages 221--242, 2003.

\bibitem{Pottmann2002}
H.~Pottmann, S.~Leopoldseder, and M.~Hofer.
\newblock {Approximation with active B-spline curves and surfaces}.
\newblock In {\em Computer Graphics and Applications, 2002. Proceedings. 10th
  Pacific Conference on}, pages 8--25, 2002.

\bibitem{Sarfraz2006494}
M.~Sarfraz, M.~Riyazuddin, and M.~H. Baig.
\newblock {Capturing planar shapes by approximating their outlines}.
\newblock {\em Journal of Computational and Applied Mathematics},
  189(1-2):494--512, 2006.

\bibitem{Saux:2003:IHI:965907.965910}
E.~Saux and M.~Daniel.
\newblock An improved hoschek intrinsic parametrization.
\newblock {\em Comput. Aided Geom. Des.}, 20:513--521, December 2003.

\bibitem{Speer1998}
T.~Speer, M.~Kuppe, and J.~Hoschek.
\newblock {Global reparametrization for curve approximation}.
\newblock {\em Computer Aided Geometric Design}, 15(9):869--877, 1998.

\bibitem{Wang2006}
W.~Wang, H.~Pottmann, and Y.~Liu.
\newblock {Fitting B-spline curves to point clouds by curvature-based squared
  distance minimization}.
\newblock {\em ACM Trans. Graph.}, 25(2):214--238, 2006.

\bibitem{xie2001automatic}
H.~Xie and H.~Qin.
\newblock {Automatic Knot Determination of NURBS for Interactive Geometric
  Design}.
\newblock In {\em Proceedings of the International Conference on Shape Modeling
  \& Applications}, page 267. IEEE Computer Society, 2001.

\end{thebibliography}

\end{document}